\documentclass[11pt]{article}
\usepackage{fullpage}
\usepackage{standalone}

\usepackage{verbatim,graphicx,  amssymb, float, fullpage, array, amsmath}
\usepackage{multirow,bigdelim}
\usepackage{array}
\usepackage{setspace}
\usepackage{caption,subcaption} 

\usepackage{color}
\usepackage[authoryear, round]{natbib}
\bibpunct{(}{)}{;}{a}{,}{,}

\graphicspath{{./__cache__/}}
\makeatletter
\def\input@path{{./__cache__/}}
\makeatother

\renewcommand{\cite}{\citet}
\newcommand{\iid}{\mathop{\sim}\limits^{iid}  }
\newcommand{\indep}{\mathop{\sim}\limits^{ind}  }
\newcommand{\dirichletd}{\text{Dirichlet}  }

\newcommand{\betad}{\text{Beta}  }
\newcommand{\multinomiald}{\text{Multinomial}}
\newcommand{\discreted}{\text{Discrete}}
\newcommand{\bx}{\mathbf{x}}
\newcommand{\by}{\mathbf{y}}
\newcommand{\bzero}{\mathbf{0}}
\newcommand{\bN}{\mathbf{N}}
\newcommand{\bn}{\mathbf{n}}
\newcommand{\cY}{\mathcal{Y}}
\newcommand{\cM}{\mathcal{M}}

\newcommand{\bpi}{\boldsymbol{\pi}}
\newcommand{\brho}{\boldsymbol{\rho}}
\newcommand{\blambda}{\boldsymbol{\lambda}}
\newcommand{\bomega}{\boldsymbol{\omega}}

\usepackage{listings}

\newcommand{\annotations}{}
\ifdefined\annotations
    \newcommand{\dmv}[1]{\textcolor{red}{\textbf{[#1 --DMV] }}}
    
    \usepackage[bordercolor=white,backgroundcolor=gray!30,linecolor=black,colorinlistoftodos]{todonotes}
    
\else
    \newcommand{\dmv}[1]{}
    
\fi

\title{Estimating the Number of Fatal Victims of the
Peruvian Internal Armed Conflict, 1980-2000: an application of modern multi-list Capture-Recapture techniques}
\date{\today}

\ifdefined\annon
\author{}
\else
\author{Daniel Manrique-Vallier\footnote{Department of Statistics, Indiana University, Bloomington, IN, USA (e-mail dmanriqu@indiana.edu)},
    Patrick Ball\footnote{Human Rights Data Analysis Group, San Francisco, CA, USA.},
    and David Sulmont\footnote{Department of Social Sciences, Pontificia Universidad Cat\'olica del Per\'u., Lima, Peru.}.
}
\fi

\begin{document}
\maketitle

\begin{abstract}
We estimate the number of fatal victims of the Peruvian internal armed conflict between 1980--2000 using stratified seven-list Capture-Recapture methods based on Dirichlet process mixtures, which we extend to accommodate incomplete stratification information. We use matched data from six sources, originally analyzed by the Peruvian Truth and Reconciliation Commission in 2003, together with a new large dataset, originally published in 2006 by the Peruvian government. We deal with missing stratification labels by developing a general framework and estimation methods based on MCMC sampling for jointly fitting generic Bayesian Capture-Recapture models and the missing labels. Through a detailed exploration driven by domain-knowledge, modeling and refining, with special precautions to avoid cherry-picking of results, we arrive to a conservative posterior estimate of
58,234
($CI_{95\%}$ = [56,741, 61,289]),
and a more liberal estimate of
65,958
($CI_{95\%}$ = [61,462, 75,387])
 fatal victims. We also determine that the Shining Path guerrillas killed more people than the Peruvian armed forces.
We additionally explore and discuss estimates based on log-linear modeling and multiple-imputation. We finish by discussing several lessons learned about the use of Capture-Recapture methods for estimating casualties in conflicts.
\end{abstract}


\section{Introduction}

After the internal armed conflict of the 1980s and 1990s, Peru chose to confront the legacy of violence with public memory by creating a Truth and Reconciliation Commission (henceforth CVR by its acronym in Spanish). Among the most important points of fact at stake in the Commission's analysis were \emph{how many people were killed} and \emph{who did it}? Specifically, of the various armed groups that participated in the conflict, what were their relative proportions of responsibility? Was the violence committed primarily by agents of the Peruvian state (EST), or by the Maoist guerrillas of the Shining Path (SLU)?

Both questions were addressed in the Commission's 2003 report \citep[][]{Ball2003}. The CVR estimated the size of the population of victims of lethal violence using stratified log-linear capture-recapture (CR) methods \citep{fienberg1972mrc} based on approximately 24,000 records previously collected by Peruvian human rights groups, by the Peruvian state, as well as the Commission itself \citep[see][]{Ball2003}. The CVR reached two important conclusions. First, that the total number of deaths was in the vicinity of 69,280. Second, that 46\% of the deaths were caused by the SLU guerrillas, 30\% by the agents of the Peruvian state, and 24\% by ``other perpetrators.''

\citet{Ball2003} study faced several challenges that led to important limitations. First, the data, while abundant, was far from optimal. Cross-tabulations according to source were highly sparse. Only the CVR's own dataset contained a significant number of reports of killings by SLU perpetrators. Capture probabilities were also highly heterogeneous. Second, the study relied on traditional log-linear modeling \citep[][Ch.6]{bishop1975dma}. Log-linear models, while versatile, have important limitations for analyzing this type of data. They have low tolerance to sparse count data, and can only model individual heterogeneity by approximating it through list-by-list interactions. Model selection is also challenging. Finally, a large portion of the records lacked perpetrator attribution. These challenges were acknowledged by the study authors, who deployed a variety of creative solutions---for example they reduced sparsity by pooling four of the six lists into a single one, resulting in a less sparse three-system data set. However these solutions were mostly ad-hoc, and many relied on untestable assumptions. We briefly review some of these issues in Section~\ref{sec:cvr:study}.

In this article, we revisit the questions that drove the CVR's study: how many Peruvians died in the internal armed conflict, and who killed them.  This study overcomes many of the CVR's original limitations. First, we have secured a new dataset that was not available in 2003.  This dataset is larger than any of the previously analyzed, and critically, it provides the needed cases to allow direct estimation of killings by SLU perpetrators.  It also provides enough data to overcome some of the sparsity limitations, allowing estimation with up to seven simultaneous lists---cf. three in \citet{Ball2003}.  Second, we apply techniques better suited for this type of data. Specifically, we extend and apply the Bayesian non-parametric latent class model (LCMCR) proposed by \cite{Manrique:LCM:2016:Capture:Recapture}, which directly models heterogeneity of capture probabilities.  This method, which was not available in 2003, has the added advantages of high tolerance for sparse tables, and of sidestepping some of the difficult issues of model section that plague other approaches.  We briefly describe the LCMCR approach in Section~\ref{sec:LCMCR}.

As part of our analysis we propose a general Bayesian strategy for dealing with missing stratification labels in a simple and principled way. We use this strategy to develop an extended version the LCMCR, and use it to deal with the missing perpetrator label problem. We describe our strategy and extension in Section~\ref{sec:LCMCR:MISS}, and present an efficient MCMC algorithm for posterior inference in the Appendix. We show the results of analysis that combines the imputation of the perpetrator stratification labels with an estimate of the missing records. We conclude that with new data and new methods, our recommended estimate is lower than but similar to the CVR's 2003 estimate . More importantly, the estimated proportions of responsibility attributed to the State and to the Maoist insurgents of the Shining Path we present here are consistent with the CVR's analysis.

\subsection{Background: The Peruvian Internal Armed Conflict (1980-2000)}
\label{sec:conflict}

The Peruvian Internal Armed Conflict sparked in a very particular context of Peruvian history. When Shining Path (\textit{Sendero Luminoso}) initiated its armed struggle on May, 1980, Peruvian society was on the eve of its first democratic elections after 12 years of military rule. The ``Revolutionary Government of the Armed Forces,'' which ruled the country between 1968--1980, undertook some of the most radical and progressive social reforms in Peruvian society. Even though social and economic inequalities remained steep, in 1980 Peru was a society undergoing major social, political and democratic transformations: a large portion of its rural and indigenous population was getting access to citizenship rights; vast amounts of land were redistributed among peasant and indigenous peoples; and radical left-wing parties and social organizations were getting access to the formal democratic process.

In that context, Shining Path, a small radical Maoist political party, decided to initiate its armed struggle to start a communist revolution in Peru. Shining Path's influence at the time was mainly limited to some provinces of the department of Ayacucho, one of the least developed regions of the country. None foresaw that this small and highly radicalized group would be able to unleash such a wave of violence in the coming years, at a time when the democratic transition was opening new channels of participation and representation for diverse sectors and social groups in Peruvian society.

The internal armed conflict in Peru spanned two decades. According to the Peruvian Truth and Reconciliation Commission (CVR), it was the bloodiest conflict in the history of the Peruvian republic. Violence of this magnitude had not occurred since the anti-Spanish indigenous rebellions of the XVIII century.

The conflict began on May 17, 1980, the day of Shining Path's first armed attack in the small town of Chuschi-Ayacucho, where they burnt the electoral materials for the general elections which took place the following day.  Most of the victims reported to the CVR died between 1982 and 1993.


By the end of the 1990's, Shining Path was defeated, both militarily and politically. The final phase of the conflict started with the capture of Abimael Guzmán, Shining Path's leader, in September 1992. In parallel, a ``repentance'' law granted immunity to militants of subversive organizations that surrendered themselves to the authorities and provided information that could lead to the incrimination of other members of their organization. After that, violent events in the country declined sharply.

\section{Capture-Recapture in Conflict Casualty Estimation}
\label{sec:CR:conclicts}
The most basic quantitative question about armed conflicts is how many people were killed. During and after any significant armed conflict, it is frequent that a variety of actors, for a variety of purposes collect data about casualties and other gross human rights violations. For example human rights advocates might collect individual cases to hold perpetrators accountable, while truth commissions might do it for historical memory preservation. These data collections are almost always incomplete and, as it is usually the case with convenience samples, non-representative of the population of deaths. Thus it is usually not possible to rely on these datasets, nor on their raw combination, to draw population-level conclusions about the magnitude of conflicts.



Capture-Recapture allows us to use data like these to estimate the total mortality \citep[see e.g.][]{Manrique:etal:2013:MSE:Conflicts,lum:etal:2013:american:statistician}.  The application of  Capture-Recapture to the analysis of violence in human populations---where it is usually known by the name Multiple Systems Estimation (MSE)---is relatively new, but by now there are several examples in the literature; for an overview, see \citet{ball_price_2019}. In Guatemala, capture-recapture estimates were found to be consistent with a United Nations commission's finding that the Army committed acts of genocide against the indigenous population during the 1980s \citep[][Vol. XII]{CEH:1999}. Capture-Recapture estimates were presented as evidence in the trial of former Yugoslav President Slobodan Milošević at the International Criminal Tribunal for the Former Yugoslavia \citep{Ball2002}, to estimate the number of killings in the 1995 massacre at Srebrenica \citep{brunborg:etal:2003:srebrenica}; the total killings in Bosnia-Herzegovina during the Civil War \citep[][]{zwierzchowski:tabeau:2010:bosnia}; killings in Timor-Leste during the 1975-1999 Indonesian occupation \citep{silva2008demography}; and killings in the Colombian \emph{departamento} of Casanare \citep[][]{lum:etal:2010:bayes:averaging,Mitchell:etal:CR:bmtrcs:2013}. An estimate of the relative crude mortality rates due to homicide by the Guatemalan Army, compared for indigenous and non-indigenous communities in three counties, was presented as expert testimony in the trial for genocide of former \emph{de facto} President General Jos\'e Efra\'in R\'ios Montt \citep{BallPrice2018}. Recently, MSE has been used to estimate the population of people held in modern slavery in the UK \citep{bales_et_al_2015} and the Netherlands \citep{cruyff2017}.

\subsection{Capture-Recapture Estimation}
\label{sec:CR:general}
Capture-recapture is a family of methods for estimating the size of a closed population based on the study of the patterns of overlapping between different lists that partially enumerate it.  Let $N$ be the unknown population size, $J$ the number of available lists, and $x_{ij}=1$ if individual $i \in\{1,...,N\}$ is in list $j \in \{1...J\}$ and $x_{ij}=0$ otherwise.  Therefore any individual in the population has an associated \textit{capture pattern}, $\bx_i=(x_{i1},...,x_{iJ})$, which indicates their presence or absence from the samples.  For example, a capture pattern $(1,1,0,0)$ indicates presence in lists one and two, and absence from lists three and four.  Under this framework, any individual with capture pattern $\bzero=(0,...,0)$ is by definition unobserved.  The objective is to estimate how many individuals in the population fall into that category.  CR methods work by fitting a joint distribution model $f(\bx|\theta)$ for the capture patterns $\bx \in \{0,1\}^J$ using the observed patterns (obtained by combining all $J$ lists), and using that fitted model to project $f(\bzero|\theta)$. This means that the quality of such estimate depends crucially on how well the model, which was fitted only using observable data, projects to the unobservable pattern $\bzero$.

Early CR methods \citep[e.g.][]{Petersen1896,Darroch1958} postulated, implicitly or explicitly, simple independence models for $f(\bx|\theta)$, thus imposing stringent requirements of independence and homogeneity to the listing processes. The development of methods for the analysis of discrete multivariate data in the 1960s allowed to relax these requirements. \citet{fienberg1972mrc} proposed the use of log-linear models and the method of conditional maximum likelihood \citep[see also][]{bishop1975dma,cormack:1989:loglinear:cr}. Log-linear modeling allows to explicitly account for interactions between listing processes, including some heterogeneity-induced forms of dependence \citep{Darroch1993}.

Most of the casualty estimation projects based on CR approaches, including the CVR study, have relied on log-linear modeling. These methods have allowed researchers to leverage multiple sources data to overcome complex listing dependency patters \citep{lum:etal:2013:american:statistician}. However, they also have important limitations. With the availability of large numbers of datasets---which provides the information needed to overcome the effects of capture dependence and heterogeneity---also comes an exponential increase of possible capture patterns and a combinatoric increase of log-linear models. This has two important consequences. First, the number of possible log-linear models can be too large to allow evaluation of all of them, even for a moderate number of lists. Second, with the growth of the number of possible patterns, at a rate $2^J$, the number of contingency table cells without elements also grows. This phenomenon, known as \textit{sparsity} in the contingency tables literature \citep{bishop1975dma}, limit the range of models where estimates exist, and can also lead to unstable estimates.

In practice these challenges are most of the times met with ad-hoc solutions. In the CVR study, where  there were originally six available lists, the authors opted for pooling four of them into a single list \citep{Ball2003}. This procedure simplified the analyses, substituting a complex---or even unfeasible at the time---6-system estimation with a simpler 3-system estimation. However, in the process the authors reduced the number of observable cells from 63 to 7, giving up a large amount of the available information, ultimately requiring stronger assumptions.

\subsection{Controlling Heterogeneity: Stratification and Modeling}
Found data of the type usually available for estimation of casualties in conflicts are usually subject to strong individual heterogeneity. Data collection is influenced by geographical constrains, which might be different for different projects; some documentation projects might focus on specific types of victims; some victims are well known in their communities while others might be relatively unknown.

The simplest way of mitigating the effects of individual capture heterogeneity is through stratification \citep{sekar:deming:1944,bishop1975dma}. The idea is to segment the population according to an observed characteristic into homogeneous (or at least more homogeneous than the general population) subgroups, and estimate each sub-group independently. In the CVR study the authors determined that the probabilities of capture varied strongly across victims of different perpetrator groups and across geographic location, thus stratified accordingly. The biggest downside of stratification is the reduction of sample sizes, which leads to sparsity problems and the reduction of inferential power.

A second, complementary, way of dealing with heterogeneity is through modeling. Models that deal with individual heterogeneity date back at least to the work of \citet{Sanathanan1972a}, and have evolved considerably in the animal abundance estimation literature; see model $M_h$ and its variants in \cite{Otis1978}. Most of these techniques represent heterogeneity through some form of latent individual-level parameter expressing ``capturability". The most common structure in these settings is the Rasch model for the probability of listing.

\subsection{The LCMCR model}\label{sec:LCMCR}
In casualty estimation it is frequent that the same individual characteristics that make some individuals more likely to appear in some lists, simultaneously make them less likely to appear in others. For example, in \citet{Ball2003} data, being a victim of SLU made individuals less likely to be listed in human rights organizations' data, while simultaneously making them more likely to be listed in the CVR database (see Section~\ref{sec:cvr:study}). This structure induces a particular form of negative dependence that makes  inappropriate the use traditional heterogeneity models from the animal estimation literature---which imply symmetric heterogeneity effects; see. e.g. $M_{th}$ models in \cite{Otis1978}.

\cite{Manrique:LCM:2016:Capture:Recapture} introduced the LCMCR (Latent Class Model Capture-Recapture) as a method for non-parametric CR estimation on multi-list heterogeneous capture settings. The technique is based on the estimation of a latent stratification scheme. Each individual is supposed to belong to one of infinitely many latent strata, even though their labels are unobserved. The model describe capture patterns through an infinite-dimensional discrete mixture of product-Bernoulli (independence) distributions,
\begin{align*}
  x_{ij} | z_i &\indep \mbox{Bernoulli}(\lambda_{jZ_{i}}), \text{ $i = 1,\dots, N$ and for $j=1,\dots,J$ } \\
  z_i &\iid \discreted(\{1,2,\dots\}, (\pi_1, \pi_2, \dots)), \mbox{ for $i = 1,\dots, N$,}
\end{align*}
paired with a stick-breaking prior \citep{Sethuraman1994} for the proportions of the total population in each stratum,
$\boldsymbol\pi = (\pi_1, \pi_2, \dots) \sim SB(\alpha)$. Here $\alpha$ has the effect of concentrating bulk of the probability mass into the first coordinates of $\boldsymbol{\pi}$ (as $\alpha$ decreases), or to spread it out to the right (as it increases). \citet{Manrique:LCM:2016:Capture:Recapture} completes the specification with diffuse priors for $N$, $\alpha$, and $\boldsymbol{\lambda}$ to allow simultaneous estimation. When all parameters are estimated simultaneously, this construction allows the automatic modulation of the complexity of the mixture by concentrating probability mass into a few mixture components as required by the complexity of the heterogeneity in the sample. \citet{Manrique:LCM:2016:Capture:Recapture} proposed an efficient MCMC algorithm  for obtaining samples from the posterior distribution and demonstrated through examples using multi-list casualties data that the method has a high tolerance for sparse tables and the capacity of learning complex heterogeneity structures.


\section{The CVR Capture-Recapture study}
\label{sec:cvr:study}

\cite{Ball2003} estimated the total number of fatal victims in the Peruvian conflict using three-way stratified log-linear capture-recapture, with data from six original sources. The six datasets were independently created by CVR itself (henceforth CVR), the Peruvian Ombudsman Office (DP), the Red Cross (CICR), and three human rights organizations (COMISEDH, CEDAP and CNDDHH). They were combined through a laborious computer-assisted record linkage procedure by an 18 person team. The resulting combined dataset collectively documents 21,951 unique fatal victims. The final global estimate was 69,280 fatal victims, attributed to the Shining Path guerrilla (SLU: 31,331), state agents (EST: 20,458) and other perpetrators (OTR: 15,967).

\begin{table}[tbp]
  \centering
\begin{tabular}{lrlrrrrr}
      & & Source & \multicolumn{1}{p{2.2cm}}{Shinning Path (SLU)} & \multicolumn{1}{p{2.2cm}}{Peruvian State (EST)} & \multicolumn{1}{p{1.5cm}}{Other (OTR)} & \multicolumn{1}{p{1.5cm}}{Unknown (NOD)} & Total\\ \cline{1-1} \cline{3-8}\multirow{6}{1.8cm}{Original \cite{Ball2003} data}  &  & CVR &  9,034 &  6,056 &   643 & 1,863 & 17,596 \\ 
   &  & DP &    198 &  3,620 &    59 &    88 &  3,965 \\ 
   & \ldelim\{{4}{0.5cm}[(*)] & CEDAP &    294 &    664 &     4 &     0 &    962 \\ 
   &  & CNDDHH &    110 &  1,865 &    19 &    98 &  2,092 \\ 
   &  & COMISEDH &     78 &  2,933 &   118 &    43 &  3,172 \\ 
   &  & CICR &     27 &     65 &     2 &   311 &    405 \\ 
   \cline{1-1} New data &  & MIMDES & 12,442 &  6,109 &   562 & 1,355 & 20,468 \\ 
   \cline{1-1}\cline{3-8} &  & TOTAL & 17,687 & 12,901 & 1,158 & 3,216 & 34,962 \\ 
  \end{tabular}

  \caption{Number of available records according to source and perpetrator. MIMDES data was not available at the time of \cite{Ball2003} study. \cite{Ball2003} grouped sources marked with (*) into a single list (``ODH"). Entries in row labeled ``TOTALS'' are counts of unique individuals and therefore do not coincide with column sums.} \label{tab:breakdown:sources}
\end{table}

Despite its abundance, these data had some peculiarities that made analysis more difficult.  First and foremost, the observed proportions of responsibility attributed to the main perpetrators (SLU and EST) in different datasets were extremely different.  As shown in Table~\ref{tab:breakdown:sources}, the CVR dataset contains substantially more records of killings attributed to SLU than to EST (i.e., 9,034 [51.3\%] $>$ 6,056 [34.4\%]), while for all other sources the proportions are reversed in a rather dramatic way.  For example COMISEDH reports 2,933 (92.5\%) deaths attributed to EST, but only 78 (2.5\%) to SLU.  Moreover, most of the records documenting killings perpetrated by SLU appeared exclusively in the CVR list (9,034 out of 9,336).  Second, there was a large number of records (2,275) for which the perpetrator was not determined.  A final minor issue was that one of the datasets (CEDAP) only contained data collected in a single region (Chungui district, in Ayacucho).


These challenges were met by \citet{Ball2003} using an array of strategies. As a first step to mitigate the effects of heterogeneity they performed all estimations stratifying by 59 custom-defined geographic regions. Such level of disaggregation, together with the large number of lists, resulted in strata with severe sparsity problems. As a compromise, \cite{Ball2003} opted for pooling the four smallest lists corresponding to non-governmental organizations (CEDAP, CNNDDHH, CICR and COMISEDH) into a single list (``ODH''; see Table~\ref{tab:breakdown:sources}), thus reducing the number of available lists from six to just three. This achieved the desired reduction of the size of the number of cross-classified sources, from 64 to 8, therefore reducing the number of empty cells per stratum. However it had the side effect of giving up all the information contained in the multivariate patterns of intersection between those lists. This strategy also implicitly introduced an untestable assumption of homogeneity-across-combined-lists.

\cite{Ball2003} also sought to stratify by perpetrator, both to produce by-perpetrator estimates, as well as means of controlling the extreme capture heterogeneity (evident in the difference between counts for SLU an EST for different sources; see Table~\ref{tab:breakdown:sources}). However, the lack of records documenting SLU killings in all datasets other than CVR made it impossible to directly produce estimates for the killings attributable to SLU. \cite{Ball2003} got around this limitation by first obtaining combined estimates of SLU and EST (pooling the records corresponding to both perpetrators within each geographic strata), and of EST alone, and then forming the estimates of SLU by computing the differences. A major problem with this approach is that it ultimately depends on the quality of the two original estimates (of EST and EST+SLU). While the existence and nature of the heterogeneity among EST records is up for discussion, the existence and severity of the heterogeneity in the  aggregate is evident (see Section~\ref{sec:cvr:study}. Thus the quality these estimates ultimately rests (untestably) on how well log-linear modeling handled the heterogeneity in the SLU+EST combined grouping.

Finally, \cite{Ball2003} handled the records with missing perpetrator labels (``Unknown'' in Table~\ref{tab:breakdown:sources}) by grouping them together with other less frequent perpetrators into a single ``Other Perpetrators" category, which was estimated separately. This strategy has two major problems. First, it creates a category in which heterogeneity is likely large and complex, thus making estimation very difficult. Second, it mishandles what is actually a missing data problem. Indeed, underlying every label ``unknown," there must be a true but unknown value, which must correspond to one of the perpetrator groups. Therefore, even if estimates were perfect, this construction made it impossible to determine which of the estimated 15,967 victims in this category were actually victims of EST or SLU. This leads to underestimation of responsibility of known perpetrators, in unknown proportions; see \citet{Zwane2007}. We discuss this problem in the context of our own strategy for handling missing data, in Section~\ref{sec:LCMCR:strat}.

\section{Data and data preparation}
\label{sec:data}

For this study we used a version of the original \cite{Ball2003} data together with a new dataset. The new dataset (henceforth MIMDES) is the result of a project by the Peruvian government, called \textit{``Censo por la Paz"} (Census for Peace). This project was aimed at creating an enumeration of the victims of the conflict independently from the CVR. While ``census" is clearly a misnomer, this was an ambitious project which produced a list with a total of 20,468 uniquely identified victims (see Table~\ref{tab:breakdown:sources}). The data collection ran in several waves between 2001 and 2006, and the results were published in a series of documents with lists of nominalized victims. Unfortunately, the agency responsible for this data collection, the \textit{Programa de Apoyo al Repoblamiento} at the Peruvian Ministry of Women and Vulnerable Populations (MIMDES) was deactivated in 2006, and its digital archive was lost. We have reconstructed this database from the digital publications (four volumes in \textit{pdf} format) using optical character recognition software, followed by manual checking and formatting. Besides its large size, a salient feature of this dataset is that it contains a large number of records of killings by SLU.

An anonymized version of the individual-level data used by \citet{Ball2003} was published in 2003 as an appendix to the electronic version of the CVR's final report.  These data, however, were insufficient for our purposes: in order to identify the matching records from the MIMDES dataset we also needed the names of the victims.  We obtained this confidential information through direct agreement with the current steward of the CVR archives, the Peruvian Ombudsman office (\textit{Defensor\'ia del Pueblo}).  This nominalized datased is very similar to the one used by \citet{Ball2003}, but includes an additional quality control step performed by the CVR, which was not ready by the time of the publication of \citet{Ball2003}. The difference between the two datasets is minimal and, if anything, the additional verification increases our confidence in the quality of the data.

\begin{figure}[tbp]
  \begin{center}
    \includegraphics[trim = 0.6in 6.5in 1.9in 1.6in, clip=true, width=0.7\textwidth]{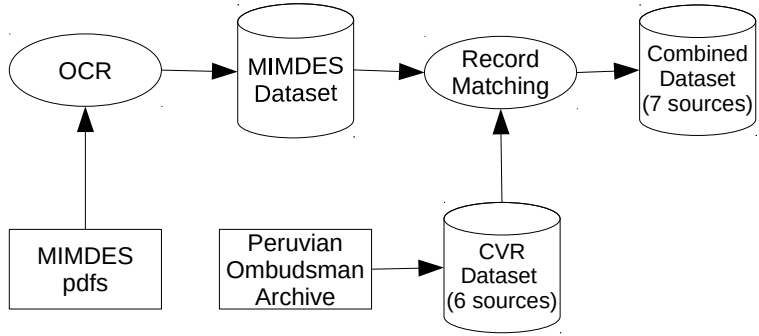}
  \end{center}
\caption{Data flow diagram of data processing.}
\label{fig:dataflow}
\end{figure}

We merged the MIMDES records with the other data (six datasets already merged by the CVR) by pairwise record linkage.  We relied primarily on the victims' names to detect matching records, and complemented with dates of birth and death and location of death to disambiguate among similar names. We sorted the data in successive passes: the first pass on the given names, the second on the paternal surname, and the third on maternal surname.  In each pass we compared records with similar names. Whenever we determined that two or more records referred to the same individual, we assigned a unique match group identifier to represent all the linked records. In a final review, we reconsidered all the match groups with more than two linked records and evaluated whether all the records actually represented the same person. If they did not, we reassigned them to smaller, more homogeneous clusters. After completing this process we found that 
619
records within MIMDES were duplicates. Figure~\ref{fig:dataflow} shows a schematic representation of the data processing.

\begin{figure}[tbp]
    \centering
    \begin{subfigure}{0.45\textwidth}
        \centering
        \includegraphics[width=\textwidth]{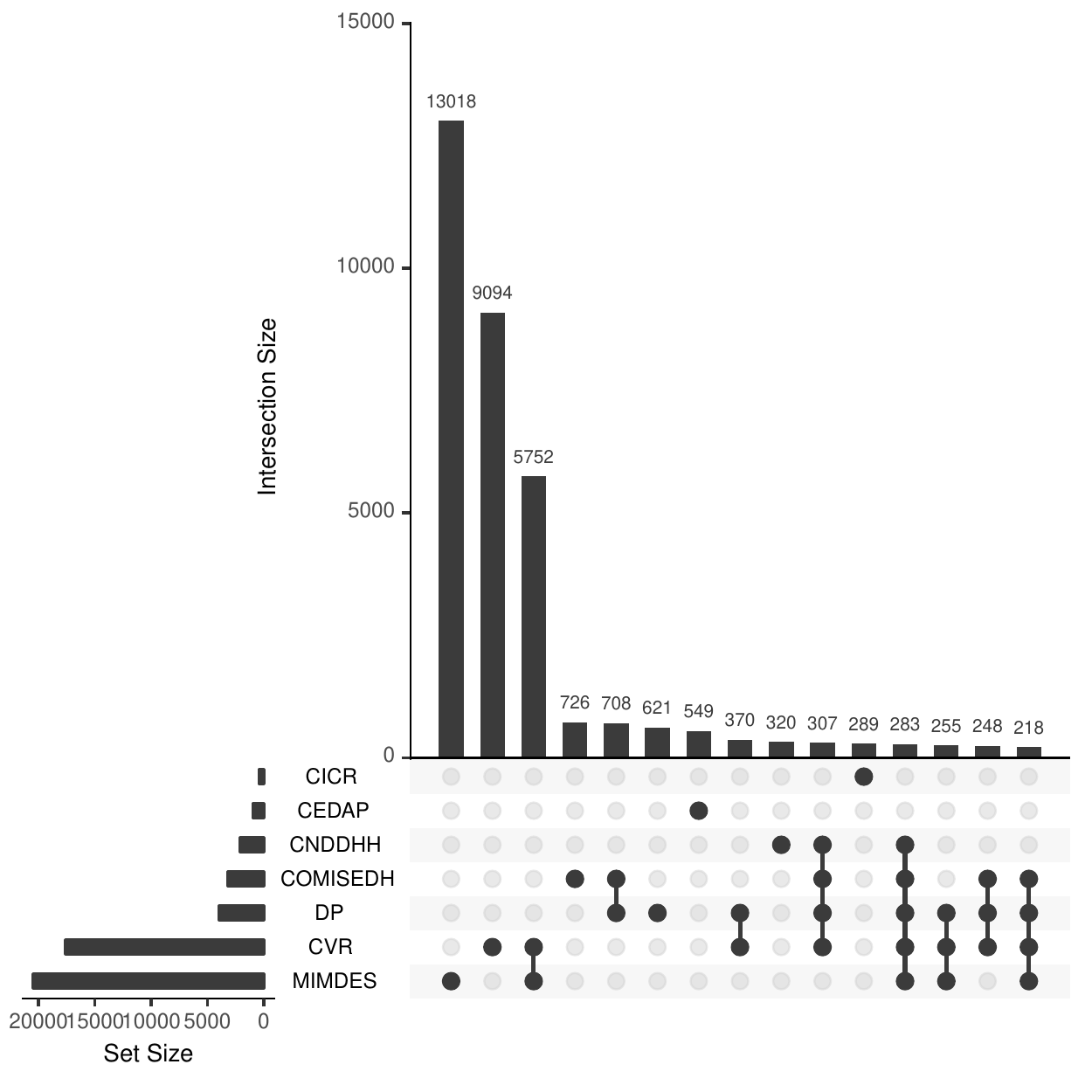}
        \caption{Aggregate}
        \label{fig:patterns:agg}
    \end{subfigure}
    \quad
    \begin{subfigure}{0.45\textwidth}
        \centering
        \includegraphics[width=\textwidth]{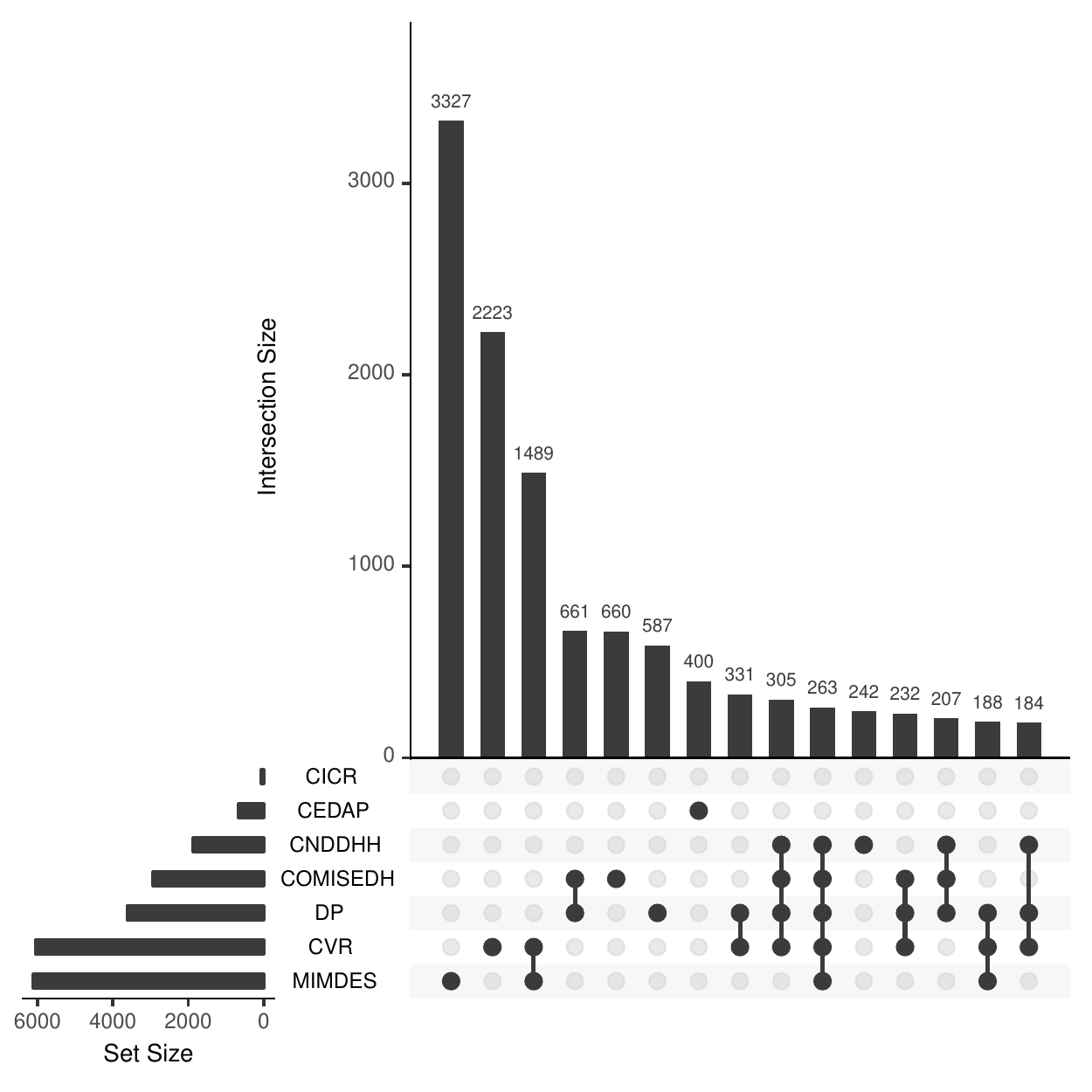}
        \caption{Peruvian State (EST)}
        \label{fig:patterns:EST}
    \end{subfigure}
    \begin{subfigure}{0.45\textwidth}
        \centering
        \includegraphics[width=\textwidth]{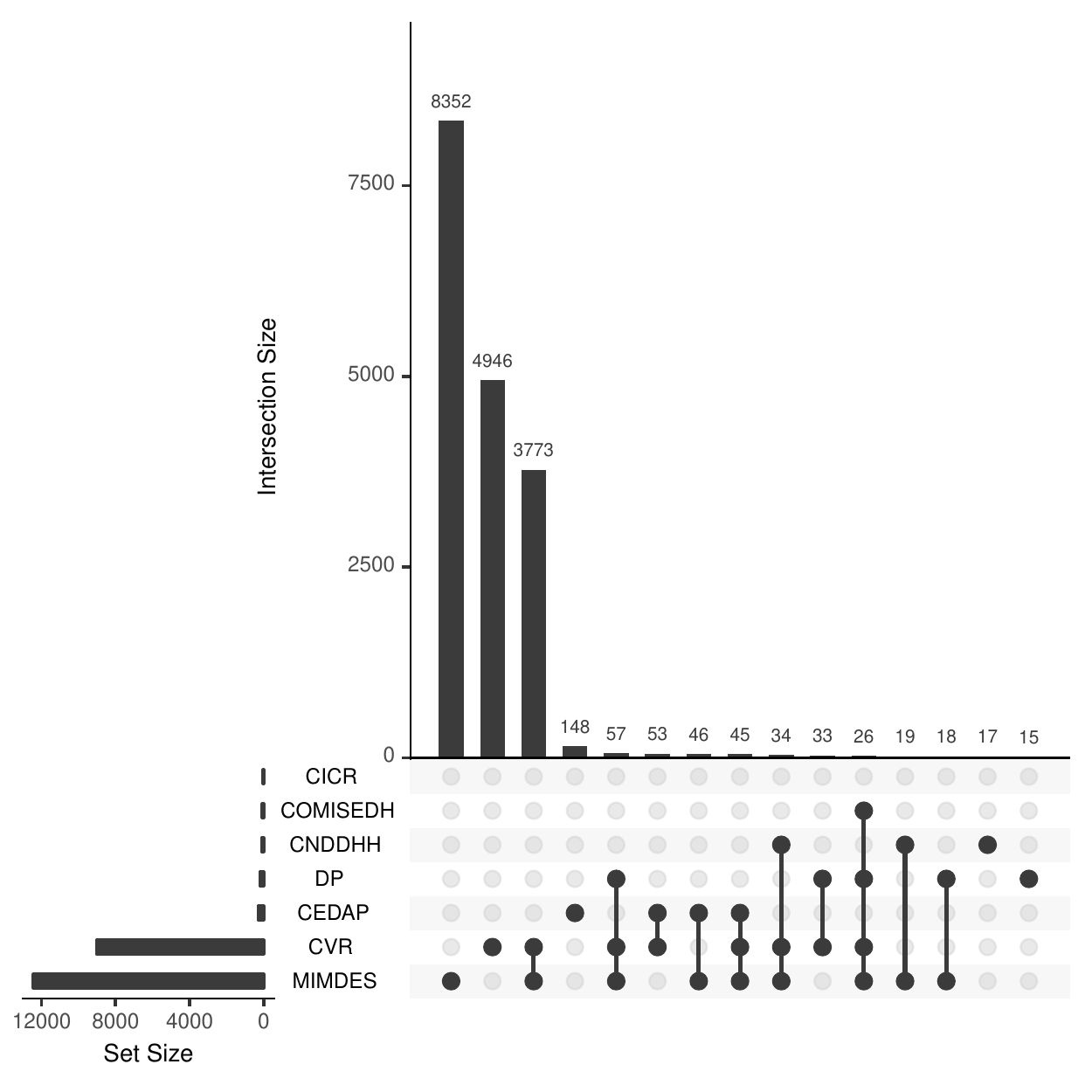}
        \caption{Shining Path (SLU)}
        \label{fig:patterns:SLU}
    \end{subfigure}
    \quad
    \begin{subfigure}{0.45\textwidth}
        \centering
        \includegraphics[width=\textwidth]{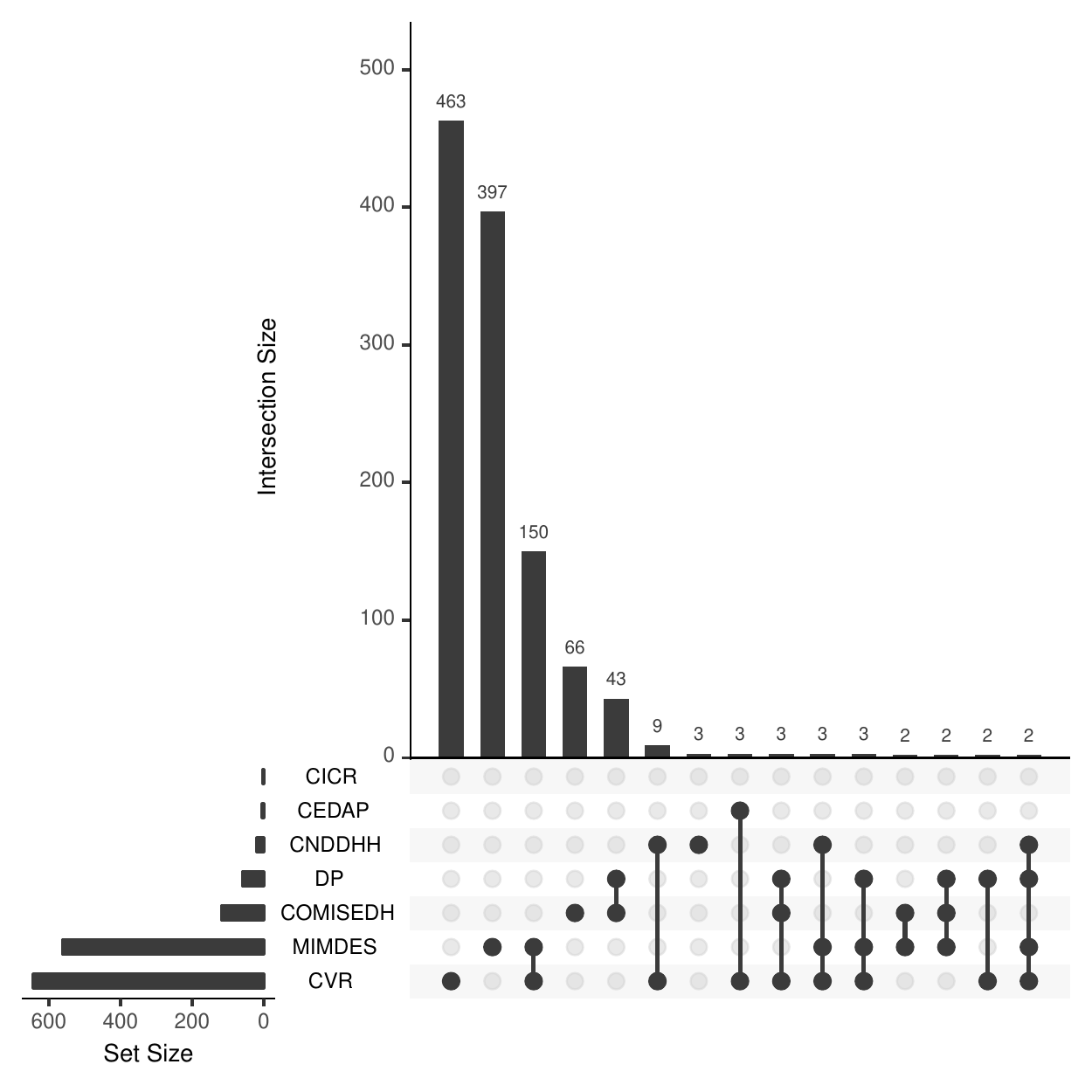}
        \caption{Other Perpetrators (OTR)}
        \label{fig:patterns:OTR}
    \end{subfigure}
    \caption{Combined dataset: the 15 most populated capture patterns by perpetrator. Plot elaborated using package UpSetR \citep{upset}}.
    \label{fig:patterns}
\end{figure}

Our final merged dataset contains a total of 34,962 unique individuals, which were identified from 49,833 original records from 7 datasets---not counting internal duplication within dataset. Figure~\ref{fig:patterns} shows the 15 most frequent capture patterns, arranged by perpetrator. As the figure shows,  the merged dataset is rich in overlaps: it contains observations from 60 of the $2^7 - 1= 127$ possible capture patterns. Even more importantly, it provides plenty of new data on killings by SLU (see panel~\ref{fig:patterns:SLU} in Figure~\ref{fig:patterns}). Indeed, the proportion of killings attributed to SLU in the MIMDES dataset is 60.8\% (12,442 out of 20,468). From these, 8,352 records are new and 4,090 were already contained in \cite{Ball2003} data.

\section{Preliminary analysis of the new combined dataset}
\label{sec:exploration}
Our objective was to produce an estimate of the total number of fatal victims, and also to apportion that total by perpetrator.  Our analysis involves a combination of exploratory data analysis, modeling, verification, and refining.  We proceeded by stages.  First, as an exploration, we produced estimates stratifying only by perpetrator, and grouping records with unknown attribution into a single class (Section~\ref{sec:byperpetrator}).  This exploration gave us a first approximation and informed us about issues with heterogeneity.  We then introduced a geographic stratification scheme with the purpose of further isolating heterogeneity (Section~\ref{sec:stratification}).  We constructed this scheme based on qualitative knowledge about the conflict, trying to keep regions as large as possible, to avoid an excessive reduction of sample sizes.  We defer a more appropriate treatment of the missing perpetrator labels until Section~\ref{sec:LCMCR:MISS}.

\subsection{First exploration: stratification by perpetrator only}
\label{sec:byperpetrator}

As discussed in Sections~\ref{sec:cvr:study}~and~\ref{sec:data}, our data exhibit extreme by-perpetrator heterogeneity.  This fact, together with our objective of producing by-perpetrator estimates, makes a by-perpetrator stratification a natural starting point. This was also the strategy followed by \citet{Ball2003}, without the MIMDES dataset. Different from them, however, we have enough data to produce estimates of the SLU stratum without resorting to indirect strategies like those described in Section~\ref{sec:cvr:study}.

In this exercise we have applied the LCMCR method described in Section~\ref{sec:LCMCR} using six lists and stratifying by perpetrator. We have excluded the CEDAP database because it only contains data from collected at a single province (Chungui, Ayacucho), so its inclusion could violate CR list-coverage assumptions. We have also grouped all the records with no known perpetrator into a single ``Undetermined'' (NOD) category. We have ran 100,000,000 iterations long MCMC chains after 500,000 burn in periods, and sub-sampled taking one sample every 10,000 samples for a total of 10,000 samples per chain. We direct interested readers to Supplement~2 for details on the computations, including summaries and detailed output.


Using this setup we obtain an aggregate estimate of $\hat N =$ 80,562. This number is larger than that of \citet{Ball2003} (69,280), and has a very large posterior dispersion ($CI_{95\%}$ = [67,470, 111,808]). These differences are mostly driven by the posterior distribution of $N_{EST}$, which is highly dispersed ($CI_{95\%}$ = [25,894, 46,552]) and also has a very large posterior median with respect to the observed count ($n_{EST} = $12,052 vs $\hat N_{EST} =$ 33,840). This was unexpected given the large observed counts, and the fact that most databases have plenty of elements for this group. We also note that the posterior distribution of $N_\text{NOD}$ is also very dispersed ($CI_{95\%}$ = [7,791, 43,241]).

\begin{figure}[tbp]
  \begin{center}
    \includegraphics[width=0.7\textwidth]{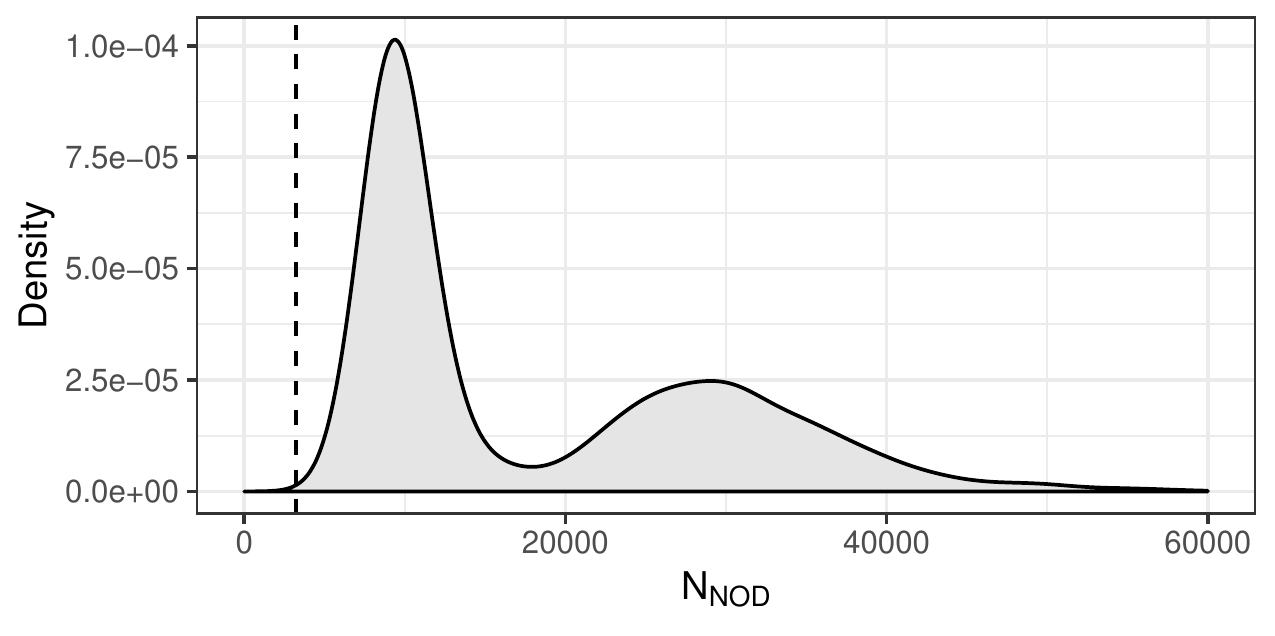}
    \caption{Posterior density of $N_\text{NOD}$ for estimation without geographic stratification. Discontinuous vertical line marks the observed count.}
    \label{fig:density:NOD:TOP}
  \end{center}
\end{figure}

These features are not necessarily an indication of problems, but they invite to take a closer look.  Inspection of the posterior density of $N_\text{NOD}$ (Figure~\ref{fig:density:NOD:TOP}; see Supplement~2 for trace plots and other output) reveals the existence of two clearly visible modes.  This feature is intriguing but hardly surprising: the NOD category does not correspond to an actual perpetrator but to records with missing perpetrator label.  Such arrangement likely groups together records that should have been distributed among the other strata, had their labels not been missing.  Therefore this category likely suffers from severe heterogeneity.  We speculate that the two modes are due to the existence of two or more underlying, more homogeneous, sub-populations, corresponding to victims of the two main perpetrators in the conflict, SLU and EST.  We also note that, while multimodality is evident, we cannot be sure about the relative weight of each mode, due to poor mixing of the chain, even after 100 million iterations; see Supplement for traceplots and other information about computations.  In the case of $N_\text{EST}$, posterior samples do not obviously hint at multimodality or poor mixing.  However, the large dispersion of the posterior distribution (considering the abundance of data) might also be an indication of heterogeneity.

\subsection{Geographic stratification and refining}
\label{sec:stratification}

Results in the previous section suggest that heterogeneity might still be severe enough to require further treatment, even after stratifying by perpetrator. In this section we add geography to the variables we use for stratification. Geographic region is a natural candidate for stratification variable. As discussed in Section~\ref{sec:conflict}, it is known that the conflict was not evenly distributed over space. It is also known that data collection efforts themselves were not geographically homogeneous---for example, CEDAP only includes data collected in a single district. Therefore, stratifying by geography, in addition to perpetrator, is likely to produce much more homogeneous sub-populations. We note that another important dimension of heterogeneity is time. However, periods of violence have been strongly geographically correlated. Indeed we note that counts of observed killings by period are strongly correlated over time. The correlations among regions over time has a median of $r=0.47$, and 95\% of the correlations are found in the range $-0.19 < r < 0.94$.

We created a suitable geographic partition by simultaneously trying to ensure both geographic contiguity and homogeneity of conflict dynamics within stratum. Our assessment of the conflict dynamics was based only on qualitative subject-matter knowledge. We still consider a NOD stratum that groups all the records with unknown perpetrator.

The hierarchy of administrative divisions in Peru comprises 25 Departments (24 plus a ``constitutional province"), 194 Provinces within Departments, and 1,828 Districts within Provinces. Our data is georeferenced at the level of district. We created our geographic stratification by dividing the Peruvian map into 12 contiguous regions following the contours of administrative divisions up to the district level, depending on the specific region. For example, we made the Chungui district a single stratum (to be able to use the CEDAP database), while we combined three departments (Cusco, Puno and Apurimac, which together comprise 33 provinces and 296 districts) to form the \texttt{SIERRA\_SUR} stratum. We sought to form strata that were as large as possible in sample size, but that still roughly correspond to the various insurgent/counter-insurgent dynamics of the conflict. We lay out the complete stratification scheme and its detailed rationale in Supplement~1.

The first part of Table~\ref{tab:stratified:noimp:perp} (labeled ``12 geo strata") shows estimates by perpetrator obtained by first computing the posterior distribution of the population size on each of the $4 \times 12 = 48$ combinations of regions and perpetrator, and then calculating the posterior distribution of the aggregate.  We also include a global aggregated estimate.  For this exercise we have used all seven lists, but not all of them in every stratum.  The reason for this is that not all sources include records for every combination of region and perpetrator.
For example, stratum \texttt{AYA\_NORTE\_CHUNGUI} has six lists available for estimating EST, but only three for estimating SLU.  Additionally we excluded lists from calculations if they 1) did not have any overlap with other lists in their corresponding stratum, and 2) listed less than four victims.

\begin{table}[tbp]
  \centering
\begin{tabular}{lrrrrrr}
  & & \multicolumn{2}{c}{12 geographic strata}& &\multicolumn{2}{c}{26 geographic strata}  \\ 
\cline{3-4} \cline{6-7}
Perpetrator & $n$ & $\hat{N}$ & $CI_{95\%}$ &   & $\hat{N}$ & $CI_{95\%}$ \\ 
  \hline
Peruvian State (EST) & 12,550 & 17,621 & [16,705, 21,645] &  & 18,079 & [17,203, 19,974] \\ 
  Shinning Path (SLU) & 17,667 & 29,900 & [29,024, 31,216] &  & 29,791 & [28,716, 31,316] \\ 
  Other (OTR) & 1,121 & 2,149 & [1,916, 2,478] &  & 1,921 & [1,587, 2,273] \\ 
  Undetermined (NOD) & 3,212 & (*)14,262 & (*)[7,241, 22,713] &  & (*)9,881 & (*)[7,301, 16,965] \\ 
   \hline
Total & 34,550 & (*)64,212 & (*)[56,634, 73,063] &  & (*)59,900 & (*)[56,518, 67,089] \\ 
  \end{tabular}

  \caption{Geographically stratified LCMCR estimates of population size aggregated by perpetrator. Estimates were computed stratifying by perpetrator and geographic region. Left side table (labeled ``12 geographic strata") corresponds to the first geographic stratification scheme, with 12 regions. Right side (labeled ``26 geographic strata") corresponds to the second one, with 26 regions. Entries marked (*) might be unreliable due to poor mixing of NOD stratum.}
  \label{tab:stratified:noimp:perp}
\end{table}

\begin{figure}[tbp]
  \centering
  \begin{subfigure}{0.45\textwidth}
    \centering
    \includegraphics[width =\textwidth]{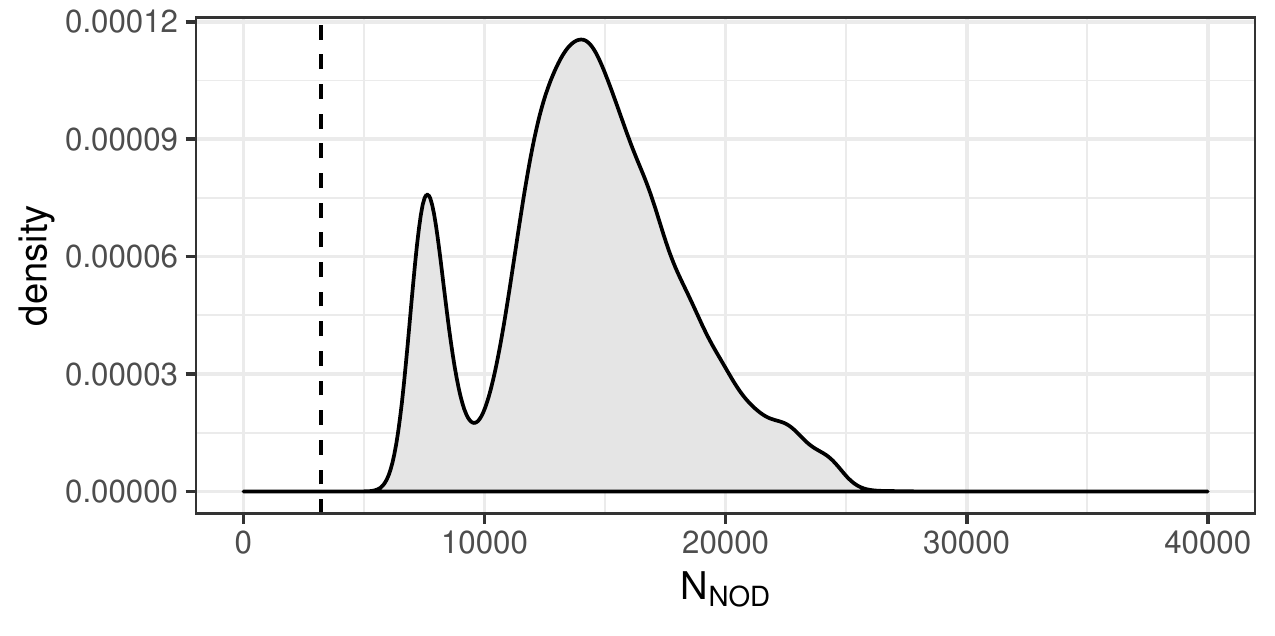}
    \caption{12-region stratification.}
    \label{fig:density:NOD:12}
  \end{subfigure}
  \quad
  \begin{subfigure}{0.45\textwidth}
    \centering
    \includegraphics[width = \textwidth]{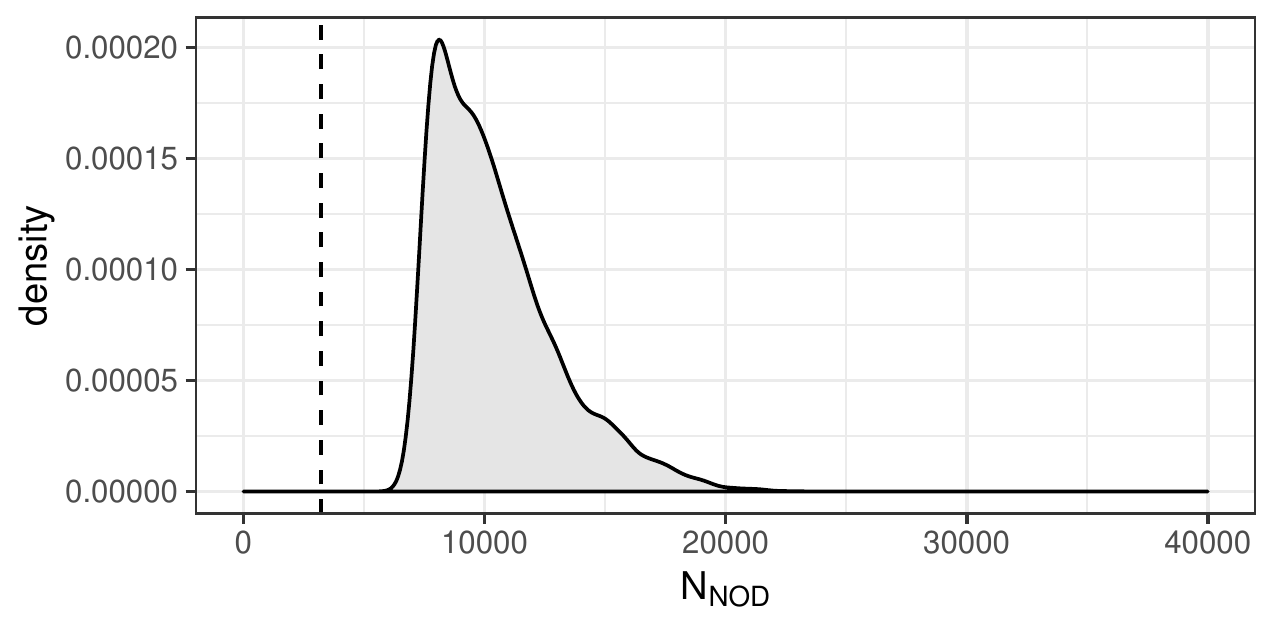}
    \caption{26-region stratification.}
    \label{fig:density:NOD:26}
  \end{subfigure}
  \caption{Posterior density of aggregated $N_\text{NOD}$ for estimation with 12- and 26-region geographic stratification. Discontinuous vertical line marks observed counts. Multimodality is evident, but relative weight of modes might be incorrect due to poor MCMC mixing.}
  \label{fig:density:NOD}
\end{figure}

The most salient result of this exercise is the marked reduction on the estimate of EST, from $\hat N_\text{EST} =$ 33,840 to $\hat N_\text{EST} =$ 17,621, and the narrowing of its corresponding credible interval to [16,705, 21,645]. These numbers are close to those in \cite{Ball2003}. The aggregated estimate for stratum NOD is roughly similar to the unstratified result, and is still very dispersed and visibly bi-modal; see Figure~\ref{fig:density:NOD:12}.  Results for SLU are very close to the unstratified ones. Taken together, these results suggest that, while there does not seem to be much geographic heterogeneity in SLU data (or that it is effectively controlled by the LCMCR latent structure), spatial heterogeneity of EST captures is large. In the case of NOD, these results reinforce the idea that heterogeneity needs to be addressed in a different way.

In order to gain a better understanding of strata's individual characteristics we have repeated this exercise using log-linear modeling \citep[][Ch.6]{bishop1975dma}. The procedure was similar to that employed by \citet{Ball2003}, but with up to seven simultaneous lists, and limited to a maximum of two-way interaction terms. We describe the full procedure and its important limitations in Supplement~1. Most of the stratum-level estimates were similar to those obtained with LCMCR. However, there were a few regions with large discrepancies between methods. Most of these discrepancies were in estimates for EST in regions that belong to the Ayacucho department. As discussed in Section~\ref{sec:conflict}, Ayacucho was the most affected area during the conflict. Therefore it is likely that heterogeneity patterns there are very complex. This hypothesis can be further supported by noting that most log-linear models in those strata are very complex, and most do not fit the data well in spite of their complexity; see Supplement~1 for detailed output. For example, stratum (\texttt{EST}, \texttt{AYA\_CENT}) ($n=$3,533, $\hat N_\text{loglin} =$7,699, $\hat N_\text{LCMCR} =$ 4,367) was estimated using 6 lists and a model with 8 two-way interaction terms, but its \textit{p-value} based on the deviance statistic is practically zero, indicating a bad model fit. Another region with similar issues is \texttt{NOR\_ORIENTE} (northeast), though there estimates differ notably not only for EST, but also for SLU and NOD.

The exercise of fitting log-linear models to these data showcases some of the important practical difficulties and limitations of log-linear capture-recapture that we discussed in Section~\ref{sec:CR:general}.  The log-linear modeling strategy consists in decomposing the joint distribution of counts in contingency tables by expressing it as a (log-linear) combination of list interactions.  This is a flexible strategy which can, in principle, represent any distribution in a contingency table, as long as we are able to include as many high-order interaction terms as needed.  In practice, however, there are strong limits to which interactions we can actually include.  Estimation of interactions between lists depend on the counts in the marginal tables corresponding to those interations \citep[see][Ch.3 and Ch.5]{bishop1975dma}.  When a multi-way margin contains zeros, its corresponding interaction term is unidentifiable and therefore not available; when the counts are small, estimates tend to be unstable.  The instability of higher order interactions is not problematic when the objective is to estimate the joint distribution of the observable cells.  In CR, however, those terms can have an inordinate impact in the projection to the unobserved cell. This problem is compounded when the dependence between lists is not caused because of true interactions between lists, but induced via individual-level traits. In these cases, it is frequent to need high order interactions to adequately model the joint distribution. This seems to be the case with many of the strata where we find very complex, but ill-fitted log-linea models, like the already mentioned (\texttt{EST}, \texttt{AYA\_CENT}).

We believe that most of the departures from independence in our data are due to individual heterogeneity. For this reason we believe that LCMCR, which directly models heterogeneity, is a better approach for this problem than log-linear Capture-Recapture. However, in order to improve our understanding of the heterogeneity patterns we decided to further explore the strata where we found large differences between the two capture-recapture approaches. We sub-divided these regions to form smaller sub-strata. The underlying assumption was that, if the problem was indeed heterogeneity, sub-stratification should ameliorate these issues. We proceeded sequentially, focusing on the five regions where the two methods exhibited the largest discrepancies, and where the dispersion was unusually large. We note that such a procedure carries the risk of altering the data distribution (a version of what is sometimes called ``researchers degrees of freedom"), due to cherry-picking of results. For this reason, at no point we based any decision about the specific way in which we sub-divided a region on the results of the computations. To enforce this principle we implemented a blinding safeguard. We split our roles: two of the authors performed the computations, not sharing the results with the third.  The third would then decide the sub-stratification based only on known dynamics of the conflict and, again, aiming at keeping geographic contiguity.

We found that, in spite of not basing the sub-stratification decisions on the results of the computations, our refining procedure had the effect of confining the differences in estimates between methods to smaller sub-regions, while ``fixing" the rest. In one (and only one) instance we abandoned our requirement of geographic continuity and grouped together the capital districts of North Ayacucho, \textit{Huanta} and \textit{San Miguel}, into a single region.  Our final geographic scheme defines 26 geographic regions, obtained from sub-dividing 5 of the original regions into 19 finer ones.  We present the detail of this scheme in Supplement~1. Interestingly,  the aggregates by perpetrator using the two stratifications are practically the same. The exception is the NOD group, which shrank slightly by approximately 4,000. We note that again, the posterior distribution of this stratum is visibly multimodal (Figure~\ref{fig:density:NOD:26}), and our estimate could be compromised by poor MCMC mixing. In the next section we address the issues with these records and their corresponding estimates.

\section{Handling Mising Data: LCMCR with incomplete stratification}
\label{sec:LCMCR:MISS}
The category NOD that we used in previous sections does not really describe any perpetrator group; instead it groups records whic are missing perpetrator attribution. This is, records labeled ``NOD" must have an underlying  true but unobserved value, which must be one of EST, SLU or OTR. Therefore, the NOD category is actually an item-level missing data indicator in the sense of \citet{Little2002}. This helps explaining the difficulties that we have experienced estimating this stratum: when taken together, records with unknown perpetrator are likely subject to the same extreme selection heterogeneity that we have observed between victims of EST and SLU, but without the possibility of differentiating between the two classes. While it is in principle possible to obtain a reasonable estimate of this group relying only on the LCMCR's handling of heterogeneity, such approach cannot help us determine how to apportion that population size between the actual perpetrators (EST, SLU and OTR). Therefore both the absolute magnitude and relative proportions attributable to the main perpetrators in the conflict will be underestimated in unknown proportions, preventing meaningful comparisons.

Partially missing stratification is a problem that has not received much attention in the CR literature so far. \citet{Zwane2007} proposed a method for attacking the problem in the context of log-linear models using the EM algorithm, however their solution does not directly translate other structures like the LCMCR. \citet{rendon2019} attempted an analysis of the original \citet{Ball2003} data using multiple imputation. However his analyses were compromised by severe statistical deficiencies \citep[see][]{manrique:ball:2019:rendon}.

In this section we propose an extension to the LCMCR to jointly handle missing perpetrator labels and population size estimation using a data-augmentation scheme. Our approach allows to estimate the joint posterior distribution of each individual's perpetrator label together with all the other parameters of interest.

\subsection{A general framework for Bayesian CR with incomplete stratification}
Following notation from Section~\ref{sec:CR:general}, let $\bx = (x_{i1},...,x_{iJ})$ be the capture pattern of individual $i = 1,...,N$ in the population. Let $y_i \in \mathcal{Y}$ be the stratum label for $i$; for example $\mathcal{Y} = \{SLU, EST, OTR\}$ in this application. We note that every individual in the \textit{population} must have an associated stratum label. This also applies to individuals which where observed ($\bx_i \neq \bzero$) but their labels were not, as well as those individuals which were not observed at all ($\bx_i = \bzero$).

Similar to the plain stratified CR we used in the previous sections, we consider that individuals from the same stratum share the same capture model. To this we now add the generation of the strata levels themselves. For $i = 1...N$ consider
\begin{equation}
  \label{eq:strata:generation}
\bx_i|y_i, \theta_y \indep f(.|\theta_{y_i}) \quad y_i | \brho \iid \discreted(\mathcal{Y}, \brho),
\end{equation}
where  $\theta_y$ is a vector of parameters specific to stratum $y \in \cY$, $f(\bx| \theta)$ is our chosen model for the capture patterns---for example the LCMCR model described in Section~\ref{sec:LCMCR}, with $\theta_y = (\blambda_y, \bpi_y, \alpha_y)$- at we have sorted the indices so that $\bx_i = \bzero$ for all $n < i \leq N$. Then labels $y_i$ for $i > n$ are by definition unobserved.  Assume that, in addition, labels $y_i$ are also unobserved when $i \in \cM \subset \{1,..,n\}$.  However, different from the previous category of missing labels, we will assume that labels in $\cM$ are \textit{missing at random} \citep[MAR,][]{Little2002}. This assumption essentally states that given the capture patterns, the missing data mechanism does not depend on the unknown value of $y_i$. Let $c_m$  be the elements of some finite set or sequence indexed by $m \in A$.  We use the notation $(c_m)_{m \in S}$ to denote the vector formed by the elements ${c_m}$ such that $m \in S \subseteq A$, following the natural ordering of $A$.  If $A$ is not itself ordered, we impose some arbitrary ordering; e.g.  lexicographic. Using this notation, let $\by_{mis} = (y_i)_{i \in \cM}$ be the vector of randomly missing labels, $\by_{obs} = (y_i)_{i \in \cM^c \cap \{i \leq n\}}$ the vector of observed labels, and $\bomega^0 = (\omega^0_y)_{y \in \cY}$ with $\omega^0_{y} = \#\{i > n : y_i = y\}$ the vector of the number of unobserved units corresponding to each level $y \in \cY$.  Finally let $\bN = (N_y)_{y \in \mathcal{Y}}$ be a vector containing all strata population sizes,  and $\bn = (n_y)_{y \in \mathcal{Y}}$ with $n_y = N_y - \omega^0_y$. Note that $n = \sum_y n_y$ and that $N = \sum_y N_y$.

In order to obtain the posterior distribution of $\bN$, we need to specify a prior distribution for $(\Theta, \bN, \brho)$. Let us for now just assume that $p(\Theta, N, \brho) = p(\brho)p(N)\prod_{y \in \cY}p(\theta_y)$. Then we have
\begin{align}
  \notag & p\left(\Theta, \bN, \brho, \by_{mis} \mid (\bx_i)_{i \leq n}, \by_{obs} \right) \\
  &\qquad \propto p(\brho)  p(N) N! \prod_{y \in \cY}
\Bigg[
p(\theta_y)I(\omega_y \geq 0) \frac{[\rho_yf(\bzero|\theta_y)]^{\omega_y}}{n_y!\omega_y!}  \prod_{i: i\leq n, y_i = y} \rho_yf(\bx_i|\theta_y)
\Bigg], \label{eq:joint:posterior}
\end{align}
from which we can estimate $\bN$ by marginalization:
\begin{equation}
  p\left(\bN \mid (\bx_i)_{i \leq n}, \by_{obs} \right)
  \propto \sum_{\by_{mis} \in \cY^{\#(\by_{mis})}} \int p\left(\Theta, \bN, \brho, \by_{mis} \mid (\bx_i)_{i \leq n}, \by_{obs} \right) d(\brho, \Theta).
\end{equation}

\subsection{Estimation: A data augmentation approach}
A Monte Carlo solution for posterior estimation of $\bN$ with missing stratification labels is to directly obtain joint samples from $p\big(\Theta, \bN, \brho, \by_{mis} \mid (\bx_i)_{i \leq n}, \by_{obs}\big)$ using \eqref{eq:joint:posterior}. Here we propose a general Gibbs sampling scheme applicable to any Bayesian  model $f(\bx_i|\theta)$ with priors of the form used in \eqref{eq:joint:posterior}. The steps we present here are generic, and they can be adapted and optimized for specific models. We present a fully worked out extension for the LCMCR in the Appendix. From \eqref{eq:joint:posterior} we derive the needed full conditional distributions to form a Gibb sampler:

\begin{enumerate}
  \item \textbf{Observed data strata imputation}: For $i \in \cM$ we have that
  \begin{align}
  p(y_i = y|...) \propto \rho_yf(\bx_i|\theta_y)
  \end{align}

  \item \textbf{Unobserved units sampling}: Instead of working with $\bN$, here it is easier to work with $\bomega^0$ and then make $N_y = \omega_y + n_y$. The full conditional distribution of $\bomega^\bzero$ is
  \begin{align}
    p(\bomega^\bzero|...) \propto p(N) N \left[(N-1)! \prod_{y \in \cY} \frac{[\rho_yf(\bzero|\theta_y)]^{\omega_y}}{\omega_y!} \right],
  \end{align}
  where the expression between square brackets is proportional to the pmf of a negative multinomial distribution. Therefore, if $p(N) \propto 1/N$ and $c_y = \rho_yf(\bzero|\theta_y)$, we have that
  $$
  \bomega|... \sim \text{NegMultinom}(n, (c_y)_{y \in \cY})
  $$

  \item \textbf{Strata proportions sampling}: The full posterior distribution is
  $$
  p(\brho|...) \propto p(\brho)\prod_{y \in \cY}\rho_y^{N_y}.
  $$
  If we take $\brho \sim \dirichletd[(\alpha_y)_{y \in cY}]$ \textit{a priori} then $\brho|... \sim \dirichletd\left[\left(\alpha_y + N_y\right)_{y \in \cY}\right]$.

  \item \textbf{Complete-data model sampling}: At this step we can condition on all the values of both $\bx_i$ and $y_i$ as if they were observed for $i=1,..N$. Thus we have for $y \in \cY$
  $$
  \theta_y|... \indep p(\theta_y)\prod_{i:y_i = y}f(\bx_i|\theta_y),
  $$
  which is equivalent to obtaining posterior samples from $\theta_y$ for each stratum independently, without considering neither missing stratification nor the truncation at $\bx_i = \bzero$.
\end{enumerate}

\subsection{LCMCR with incomplete stratification}
\label{sec:LCMCR:strat}
\newcommand{\balpha}{\boldsymbol{\alpha}}
\newcommand{\bz}{\mathbf{z}}
Here we extend the LCMCR model from \citet{Manrique:LCM:2016:Capture:Recapture} for accounting for incomplete stratification. Combining the LCMCR generative model described in Section~\ref{sec:LCMCR} and the extension from Section~\ref{sec:LCMCR:MISS}, our full complete-data generation model is
\begin{align}
  \label{eq:LCMCR:strat}
  x_{ij} | z_i, y_i &\indep \mbox{Bernoulli}(\lambda_{yjz_{i}}), \text{ $i = 1,\dots, N$ and for $j=1,\dots,J$ } \\
  z_i &\sim \discreted(\{1,2,\dots\}, (\pi_{y1}, \pi_{y2}, \dots)), \mbox{ for $i = 1,\dots, N$,}\\
  y_i &\sim \discreted(\cY, \brho), \mbox{ for $i = 1,\dots, N$,}
\end{align}
which we complete into a full Bayes specification with the independent priors
\begin{align}
  \lambda_{yjk} &\iid Beta(p_y, q_y) &\text{ for } y \in \cY, j=1..J, \text{ and } k =1,2,...K^*\\
  \bpi^y = (\pi_{y1}, \pi_{y},..., K^*) &\sim \text{SB}_{K^*}(\alpha_y) &\text{ for } y \in \cY\\
  \alpha_y &\iid Gamma(a,b) &\text{ for } y \in \cY\\
  \brho &\sim \dirichletd(1,...,1).&
  \label{eq:LCMCR:strat:priors}
\end{align}
Here SB$_{K^*}(\alpha)$ is the finite stick-breaking process with $K^*$ components, as described by \citet{ishwaranjames} as an almost-sure approximation to the infinite dimensional stick-breaking process \citep{Sethuraman1994}. We take \citet{Manrique:LCM:2016:Capture:Recapture} recommendation of taking $a=b=0.25$ as a diffuse prior specification for the concentration parameter of the stick-breaking process. In the case of $\lambda_{yjk}$, the
conditional (on latent class $k$ and stratum $k$) Bernoulli probability of appearing in list $j$, in Section~\ref{sec:full} we will discuss two alternatives, one diffuse and another more conservative.

We adapt this complete-data specification into a Capture-Recapture model with missing stratification by assuming that any individual with $\bx_i = 0$ cannot be observed, and that stratification labels for any individual with $i \in \cM \subset\{1,...,n\}$ are missing at random. A fully developed efficient MCMC sampler for this model based on the ideas from the previous section can be found in the Appendix. 

\section{Full extended LCMCR analysis of the new combined dataset}

\label{sec:full}
We have applied the extended LCMCR model outlined in the previous section to the new combined dataset. We have used the 26-region geographic partition scheme defined in Section~\ref{sec:LCMCR:strat}. Within each of the 26 regions we have jointly fitted all three perpetrators, EST, SLU and OTR, using the extended LCMCR model---records previously labeled `NOD' are now considered as belonging to the $\cM$ set, and their labels treated as MAR missing data.

We have applied the MCMC sampler described in the Appendix. Similar to our computations using the plain LCMCR model in previous sections, we have ran the chains for a total of 100 million iterations, after a burn-in period of 500,000 iterations. In order to keep the output manageable, we have also sub-sampled the chains taking one iteration for every 10,000, for a total of 10,000 samples per region. Computations took approximately 4 hours for each of the 26 geographic regions, thus requiring approximately 100 processor-hours to complete.

Computing models that simultaneously fit strata of very different sizes presents some interesting challenges. As described in Section~\ref{sec:stratification}, not all 7 lists cover all combinations of region and perpetrator. For example, in stratum \texttt{SIERRA\_CENTRO\_JUNIN} there are only
  2
lists that contain at least one observation for perpetrator OTR, while there are
  6
for EST, and
  5
for SLU. This means that for several combinations of regions and perpetrators, entire lists will contain no data. To avoid problems we tuned models so that, for each perpetrator, models only rely on lists that have at least four observations. In the few cases in which we had strata with less than two lists satisfying this condition, we have refrained from modeling the corresponding perpetrators. In those cases we kept population sizes at their observed value. This strategy has the potential of introducing a downward bias. However, since this only happens in a few very small strata (mostly OTR) to begin with, this bias must be negligible.

Another important consideration when performing simultaneous estimation is the effect of the prior specification of the latent class capturability parameters, $\lambda_{yjk}$. Remember from Section~\ref{sec:LCMCR} and Section~\ref{sec:LCMCR:strat} that these parameters represent the probability that an individual, victim of group $y$, which belongs to latent class $k$, will be listed in list $j$. We originally specified uniform priors for these parameters, expecting to obtain most of the information from the data.  This worked well for most regions.  However, in a few regions (\texttt{SELVA}, \texttt{NOR\_ORIENTE\_HUANUCO} and \texttt{NOR\_ORIENTE\_SAN\_MARTIN}), this specification led to unstable chains, where $N$ grew uncontrollably. Inspection of the computations at the level of the combination of region, perpetrator and latent class, revealed that this happened in partitions where many lists did not have elements in common. This caused posterior inference to be dominated by the (flat) prior, thus leading to very dispersed posteriors for $N$. We note that this behavior seems to depend on the configuration of capture patterns, and not on the absolute sample size of the regions---in fact two of the regions where this happened had relatively large sample sizes. We have dealt with this problem by making the prior for $\lambda_{yjk}$ weakly informative towards higher probabilities of listing, by making $a>1$ in $\lambda_{yjk} \sim Beta(a, 1)$. We note that this problem is related to the unbounded estimation risk problem identified in \citet{Johndrow:Lum:Manrique:2016:CR:Heterogeneity}.

We constructed two sets of prior distributions.  For the first one (``diffuse'') we set $a=1$ (uniform), except for \texttt{NOR\_ORIENTE\_HUANUCO}, \texttt{NOR\_ORIENTE\_SAN\_MARTIN} ($a=$ 1.1 for both), and \texttt{SELVA} ($a =$ 3).  The second prior specification (``conservative'') specifies a weakly informative prior of $\betad(2,1)$ for all regions, except for \texttt{SELVA}, for which it specifies $\betad(3,1)$.  This is an informative prior distribution, in the sense that gives a small preference to higher values of $\lambda_{yjk}$---with a prior expected value of $2/3$ and $3/4$ respectively. However, it is only mildly so, as it is only places placing the equivalent of a prior sample size of 4 units to \texttt{SELVA} and of 3 units for the rest \citep[][p39]{hoff:2009:bayes:book}.  The special treatment for the \texttt{SELVA} region was necessary to keep the posterior probability of population size under control. We note, however, that this is a small stratum which is unlikely to account for much of the population size.


\begin{table}
\centering
  \begin{subtable}{0.7\textwidth}
\begin{tabular}{lrrrr}
 Perpetrator & $n$ & $\hat n_\text{IMP}$ & $\hat{N}$ & $CI_{95\%}$ \\ 
  \hline
Shining Path (SLU) & 17,667 & 1,906 (27.9) & 33,736 & [32,879, 34,707] \\ 
  Peruvian State (EST) & 12,452 & 1,082 (27.0) & 22,331 & [21,223, 25,357] \\ 
  Other (OTR) & 1,121 & 185 (13.2) & 2,099 & [1,892, 2,362] \\ 
   \hline
Total & 31,240 & 3,173 & 58,234 & [56,741, 61,289] \\ 
  \end{tabular}

    \label{tab:LCMCR:mis:lib}
    \caption{Conservative prior.}
  \end{subtable}
  \begin{subtable}{0.7\textwidth}
\begin{tabular}{lrrrr}
 Perpetrator & $n$ & $\hat n_\text{IMP}$ & $\hat{N}$ & $CI_{95\%}$ \\ 
  \hline
Shining Path (SLU) & 17,667 & 1,887 (27.1) & 34,267 & [33,048, 36,095] \\ 
  Peruvian State (EST) & 12,452 & 1,098 (26.0) & 29,391 & [25,114, 38,400] \\ 
  Other (OTR) & 1,121 & 188 (13.4) & 2,232 & [1,802, 2,618] \\ 
   \hline
Total & 31,240 & 3,173 & 65,958 & [61,462, 75,387] \\ 
  \end{tabular}

    \label{tab:LCMCR:mis:cons}
    \caption{Diffuse prior.}
  \end{subtable}
  \caption{Global and by-perpetrator estimates using extended LCMCR for two prior specifications. Estimates were computed jointly estimating all perpetrators (including with missing labels) for each geographic region (26-region scheme). Point estimates are posterior medians. Column $\hat n_\text{IMP}$ contains estimates of the number of records from NOD that correspond to each perpetrator with posterior standard deviations between parentheses; its total coincides with the total of records originally labeled ``NOD".}
  \label{tab:LCMCR:mis}
\end{table}

\begin{figure}
  \centering
  \begin{subfigure}{0.45\textwidth}
    \centering
      \includegraphics[width=\textwidth]{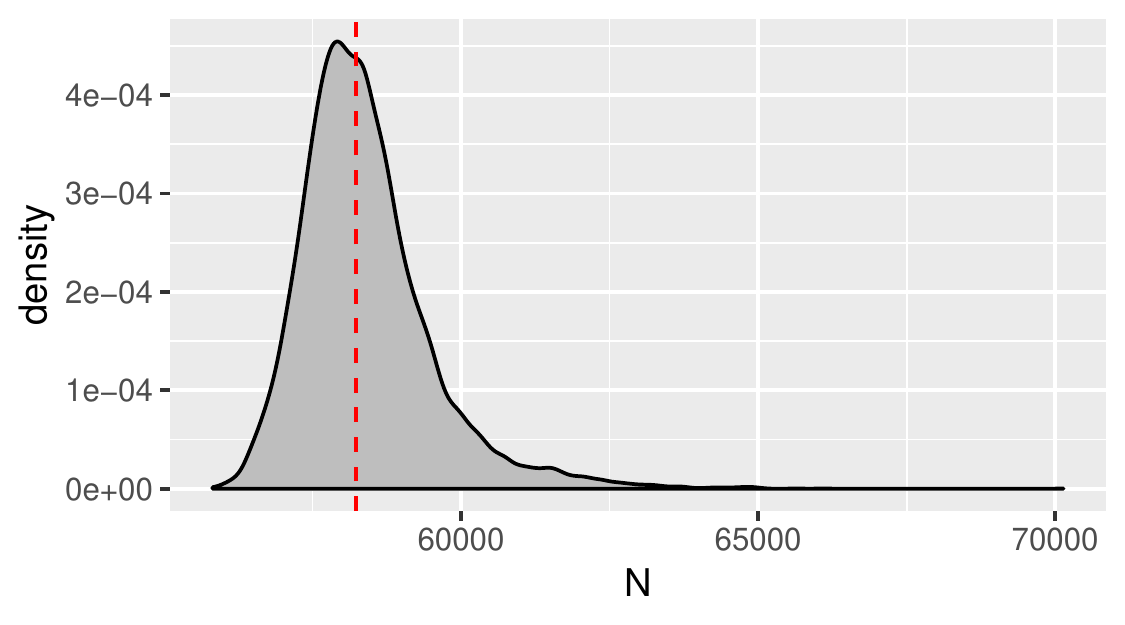}
      \label{fig:LCMCR:mis:lib}
      \caption{Conservative prior}
  \end{subfigure}
  \quad
  \begin{subfigure}{0.45\textwidth}
    \centering
      \includegraphics[width=\textwidth]{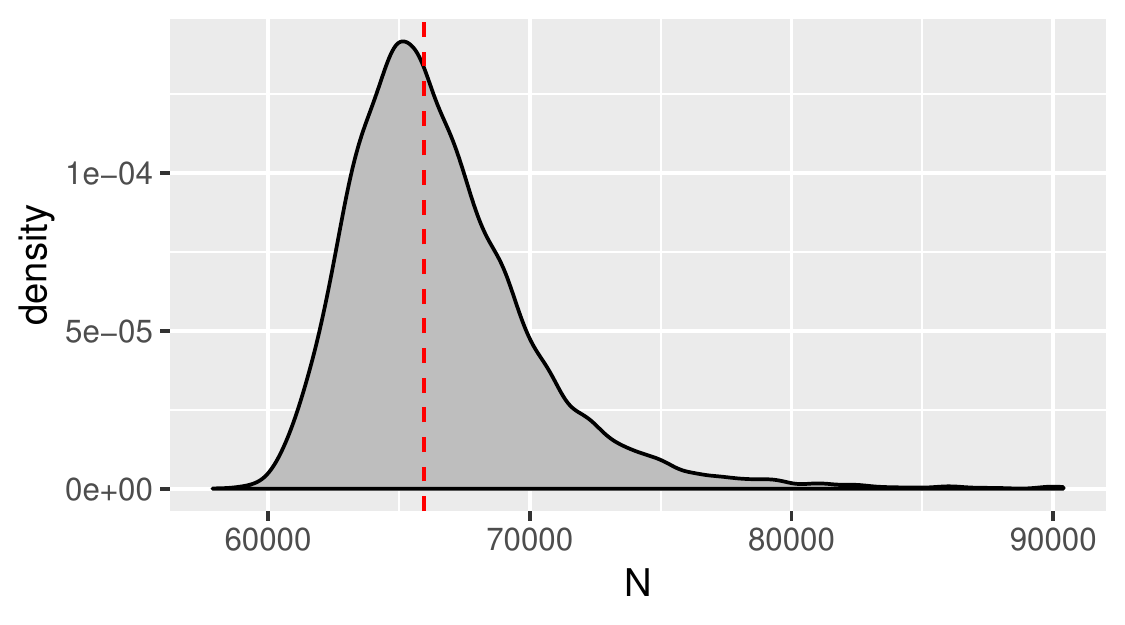}
      \label{fig:LCMCR:mis:lib}
      \caption{Diffuse prior}
  \end{subfigure}
  \caption{Posterior density of total population size using extended LCMCR model for two prior specifications. Discontinuous vertical line marks the posterior median.}
  \label{fig:LCMCR:mis}
\end{figure}

Table~\ref{tab:LCMCR:mis} shows region-aggregated extended LCMCR estimates using both prior schema. These tables do not include the label ``NOD'', as we have now treated it as a missing data indicator, and properly allocated its members to the other three categories as part of the posterior inference procedure. We show the posterior average number of individuals imputed per category in the column labeled ``$\hat n_\text{IMP}$". As expected, estimates under the diffuse prior are slightly larger and more dispersed, with $\hat N =$
65,958
($CI_{95\%}$ = [61,462, 75,387]),
than those under the mildly informative conservative priors, at $\hat N = $
58,234
($CI_{95\%}$ = [56,741, 61,289]).
Figure~\ref{fig:LCMCR:mis} shows kernel density estimates these posterior distributions.

\begin{figure}
  \centering
  \begin{subfigure}{0.4\textwidth}
    \includegraphics[width=\textwidth]{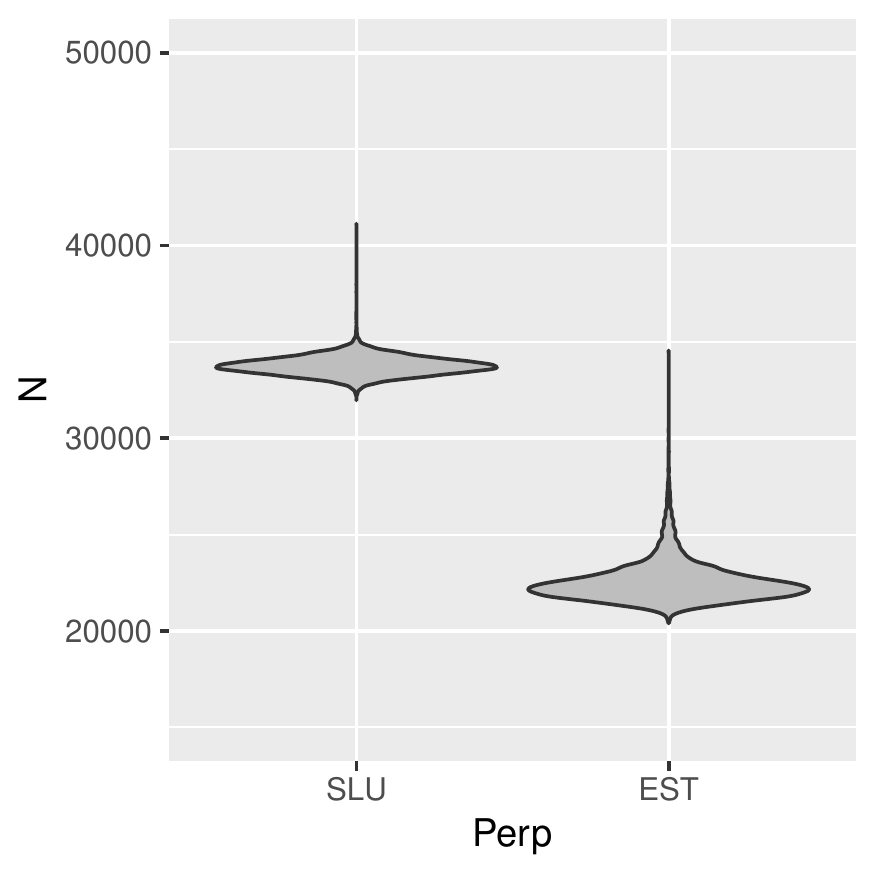}
      \caption{Conservative prior}
      \label{fig:LCMCR:mis:violin:const}
  \end{subfigure}
  \quad
  \begin{subfigure}{0.4\textwidth}
    \includegraphics[width=\textwidth]{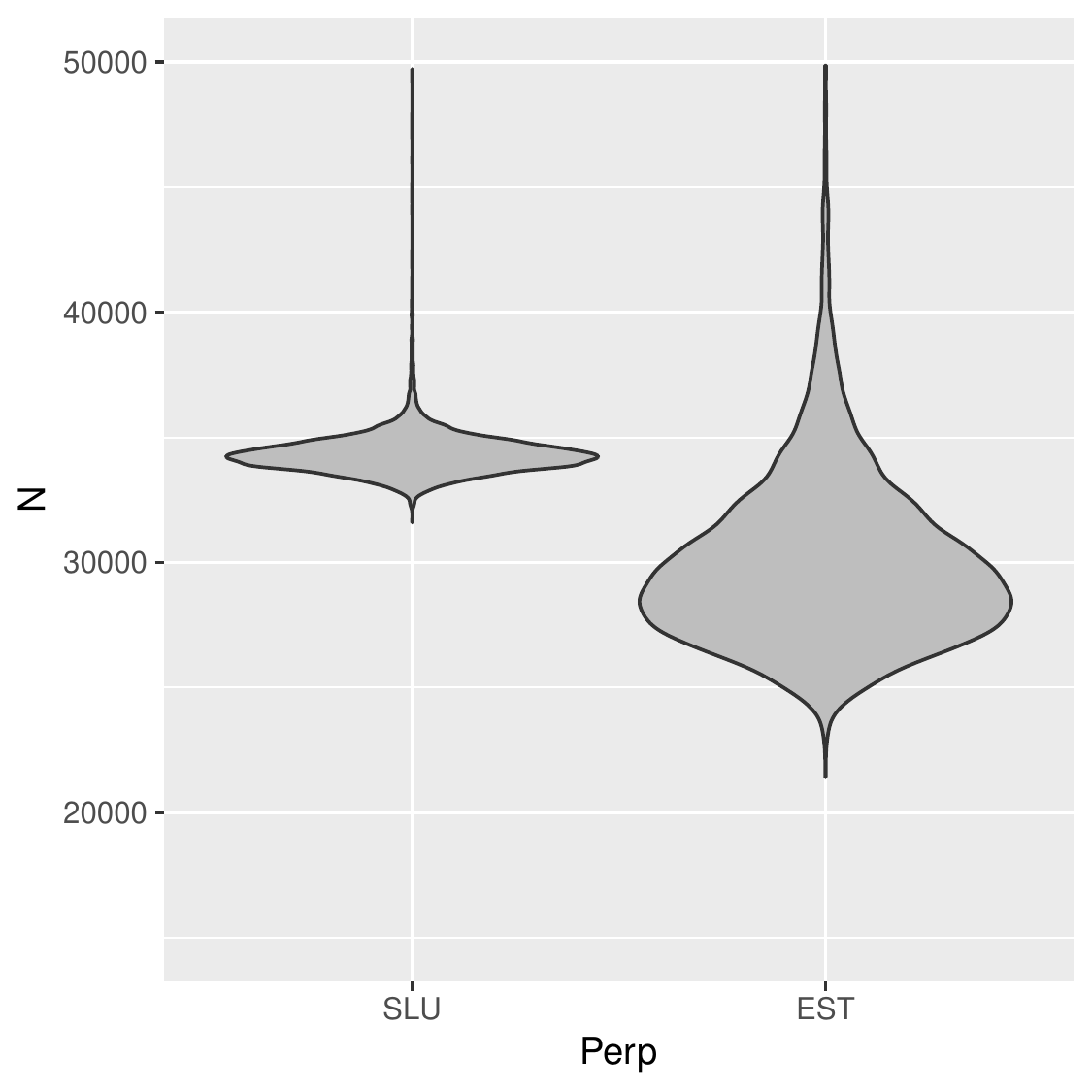}
      \caption{Diffuse prior}
      \label{fig:LCMCR:mis:violin:lib}
  \end{subfigure}
  \caption{Posterior density of population size for perpetrators EST an SLU using extended LCMCR model for two prior specifications.}
      \label{fig:LCMCR:mis:violin}
\end{figure}

The difference in global estimates between the diffuse and the conservative prior specifications is mostly driven by the posterior distribution of EST. Indeed, for the diffuse specification we have $\hat N_\text{EST} =$
29,391
($CI_{95\%}$ = [25,114, 38,400]),
while for the conservative specification we have $\hat N_\text{EST} =$
22,331
($CI_{95\%}$ = [21,223, 25,357]).
Figure~\ref{fig:LCMCR:mis:violin} shows a graphical representation of these posterior densities, as well as those corresponding to SLU. We can see that the posterior distribution for EST under conservative priors (Figure~\ref{fig:LCMCR:mis:violin}, panel~\ref{fig:LCMCR:mis:violin:const}) is much more concentrated, and is centered around a smaller value than the corresponding one for diffuse priors (Figure~\ref{fig:LCMCR:mis:violin}, panel~\ref{fig:LCMCR:mis:violin:const}). On the other hand, estimates for SLU are similar between the two specifications, with both estimates around 34,000 and a similar dispersion.

The global estimate under the diffuse specification ($\hat N = $ 65,958) is similar to the estimate obtained without missing data handling, in section~\ref{sec:stratification}. By-perpetrator results are, unsurprisingly, larger, reflecting the addition of the previously missed NOD individuals. These results suggest that treating records with missing attribution as a separate group can in fact allow to reasonably estimate the aggregate total, although it does not allow to correctly apportion that total among perpetrators. We also note that  the global total is still remarkably similar to that of \cite{Ball2003} ($\hat N = 69,280$).

\begin{figure}
  \centering
  \begin{subfigure}{0.4\textwidth}
    \includegraphics[width=\textwidth]{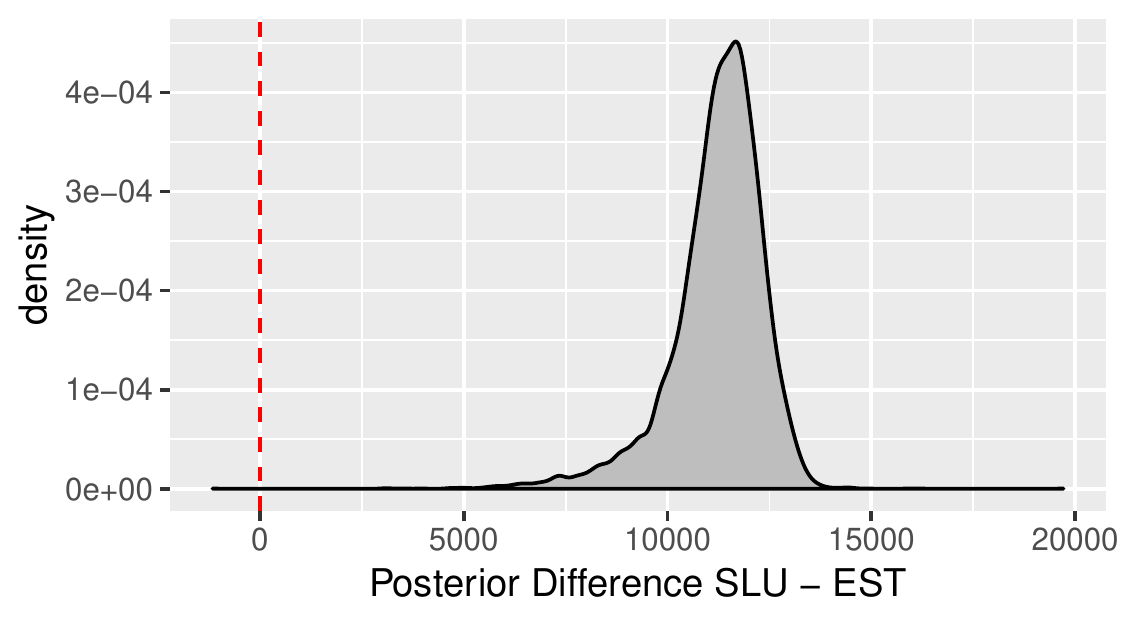}
      \caption{Conservative prior}
      \label{fig:LCMCR:mis:diff:cons}
  \end{subfigure}
  \quad
  \begin{subfigure}{0.4\textwidth}
    \includegraphics[width=\textwidth]{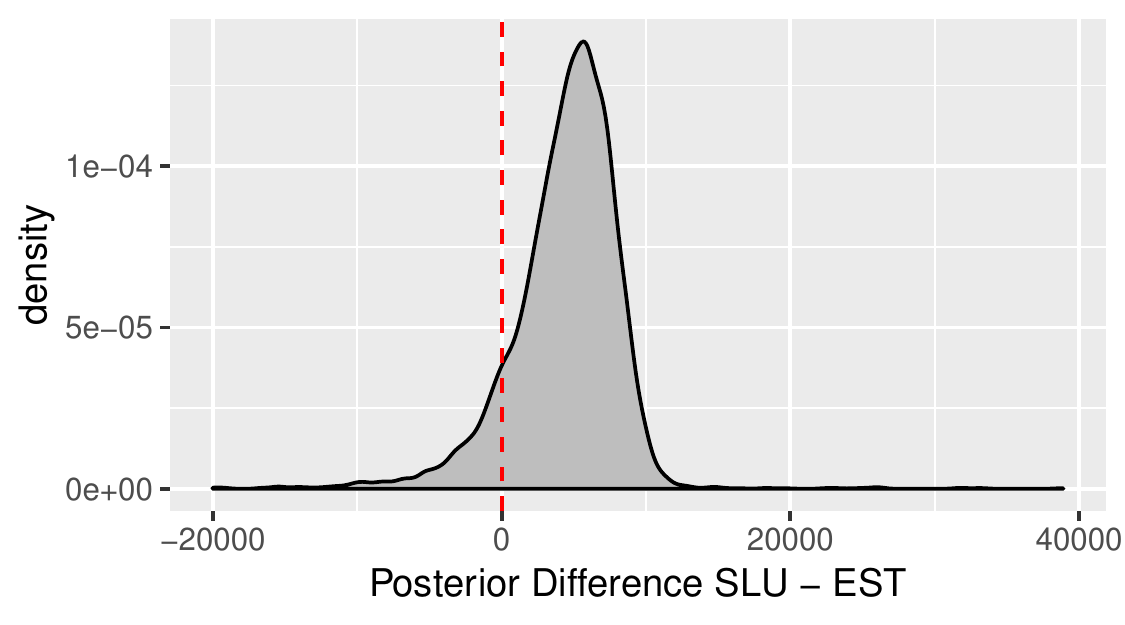}
      \caption{Diffuse prior}
      \label{fig:LCMCR:mis:diff:lib}
  \end{subfigure}
  \caption{Posterior density of difference $N_{SLU} - N_{EST}$ for two prior specifications.}
  \label{fig:LCMCR:mis:diff}
\end{figure}

We also looked at the difference in the attribution of responsibility of SLU and EST. This was an important conclusion of the CVR: the Shining path killed more people than agents of the Peruvian state. For this we calculated the posterior distribution of $N_{SLU}-N_{EST}$. Figure~\ref{fig:LCMCR:mis:diff} shows the posterior densities of this parameter under our two prior specifications. From panel~\ref{fig:LCMCR:mis:diff:cons} of Figure~\ref{fig:LCMCR:mis:diff} it is clear that under the conservative prior the posterior probability $\Pr(N_\text{SLU} > N_\text{EST}|\text{Data}) \approx 100\%$. Furthermore, we calculate a 95\% posterior probability that the difference is greater than
9,070.
Under the diffuse prior the posterior uncertainty associated with these hypotheses is larger due to the extra dispersion of $N_\text{EST}$. Nonetheless, we calculate $\Pr(N_\text{SLU} > N_\text{EST}|\text{Data}) \approx$
89.7\%.
Therefore the CVR's conclusion about the relative difference in responsibility attributed to the EST and SLU holds under our analysis.

We repeated these calculations using plain LCMCR and log-linear models, handling missing data using Multiple Imputation \citep{rubin:1987}. We describe this analysis in Supplement~1, and present detailed output in Supplement~2. LCMCR point estimates were very similar to those obtained through the joint procedure, although posterior dispersion estimates where somewhat different. The disagreement between measures of dispersion is likely due to the violation of the normality assumption underlying multiple imputation techniques. Log-linear estimates were higher, mostly driven by a higher estimate for $N_\text{EST}$. We note, however, that log-linear estimation was severely limited by poor model fitting and sparsity. We direct interested readers to Supplement~1 for details.

\section{Discussion}\label{sec:discussion}



In this article we revisit the statistical findings that the Peruvian Truth and Reconciliation Commission published in 2003, overcoming many of its limitations by adding considerably more data, and developing and using new methods. Using an extended version of the nonparametric latent class CR method from \cite{Manrique:LCM:2016:Capture:Recapture} we have produced two sets of estimates. Our recommended one, obtained using a conservative prior specification, states that
58,234 
($CI_{95\%}$ = [56,741, 61,289]) 
people were killed or disappeared between the years 1980 and 2000 as a consequence of the Peruvian internal armed conflict. From these, we attribute
22,331 
($CI_{95\%}$ = [21,223, 25,357]) 
to the Peruvian state (EST) and
34,267 
($CI_{95\%}$ = [33,048, 36,095]) 
to the Shining Path guerrillas (SLU). The rest correspond to killings perpetrataed by other agents. Our second set of results, obtained through a more diffuse specification, states that
65,958 
($CI_{95\%}$ = [61,462, 75,387])
people were killed. The difference is mostly driven by a larger estimate for the Peruvian State, which increased to
29,391 
($CI_{95\%}$ = [25,114, 38,400]). 

These results are different from, but consistent with \citet{Ball2003}. An important driver of the differences is our treatment of missing record labels.  \cite{Ball2003} treated unspecified perpetrators as a different category (NOD) from the main perpetrators. This arrangement induces a partition of \textit{population} into non-overlapping strata: victims of unidentifiable perpetrators, victims of Shinning Path for which it is possible to identify the perpetrator, and victims of the Peruvian State forces for which it is possible to identify the perpetrator. Therefore, as long as these strata are sufficiently homogeneous, it should be possible to obtain an estimate the global total.  In fact, our aggregate result using the diffuse prior specification,
$\hat N = $
65,958 
($CI_{95\%}$ = [61,462, 75,387])  
is very similar to \citet{Ball2003}
($\hat N =$ 69,280, $CI_{95\%}$ = [61,007, 77,551]). 
This coincidence is remarkable, considering that we have used different methods and added a whole new dataset which is larger than all of the previous ones. Nonetheless, the 2003 approach has two problems. First, the NOD category is an unknown mixture of the victims of the two main perpetrator groups for which the only common trait is that it is not possible to determine the perpetrator. Therefore heterogeneity in that stratum is likely to be high and difficult to estimate. This became evident in all the difficulties that we had trying to estimate the NOD stratum in Section~\ref{sec:exploration}. Second, as \cite{Zwane2007} warn, estimates corresponding to EST and SLU are restricted to the subset of the population \textit{for which it is possible to identify the perpetrator}. Thus, estimates in those categories underestimate (in unknown proportions) the true death toll due to the main perpetrators.  Estimates for SLU and EST under the diffuse prior are larger than those by \citet{Ball2003}, as should be expected because we are including our estimates of the apportioning of the records without perpetrator attribution.

Our treatment of missing perpetrator labels avoids all these pitfalls. Following advice from the discussion of \cite{Manrique:LCM:2016:Capture:Recapture}, we have developed a new general framework for dealing with missing stratification labels in Bayesian CR models. Using this framework we have extended the LCMCR, and developed a corresponding estimation algorithm. Instead of considering an artificial 'NOD' group for records with missing labels, our method consists in jointly estimate the missing label assignments together with the rest of quantities of interest.

Our recommended global estimate,
58,234,
($CI_{95\%}$ = [56,741, 61,289]),
is lower than the 2003 total \citep[cf.][{$\hat N =$ 69,280, $CI_{95\%}$ = [61,007, 77,551]}]{Ball2003}. However, our conclusions regarding the relative proportions by perpetrator are consistent with the 2003 estimates.  This was perhaps the most historically important of the Truth Commission's statistical findings: the guerrillas committed considerably more killings than the Peruvian state. This finding distinguishes the Peruvian conflict from other Latin American internal conflicts such as Guatemala, El Salvador, or Argentina in which state agents were responsible for twenty or thirty times more violence than their guerrilla adversaries. Employing our Bayesian approach allowed us to directly investigate the posterior distribution of $N_\text{SLU} - N_\text{EST}$ and to make probabilistic statements about the magnitude of the difference. Using our recommended conservative prior, we estimate that the posterior probability that SLU has killed more people than EST is practically 100\%. Moreover, there is a 95\% probability that this difference is greater than
9,070. Our first conclusion still holds under the diffuse prior, although not as starkly, with $\Pr(N_\text{SLU} > N_\text{EST} | Data) \approx$
89.7\%. 
We also performed an alternative analysis using plain LCMCR, handling the missing data labels through multiple imputation. Results were similar to those obtained under the extended LCMCR with the diffuse prior specification.

Throughout our analysis we also applied log-linear CR using seven lists. This was different from \citet{Ball2003} in that we did not combine the smaller lists into a single one, and we did not employ the ad-hoc subtraction method to make up for the scarcity of SLU data. Global results were different from those obtained using LCMCR. We noted that the differences between log-linear and LCMCR were almost exclusively concentrated on a few strata where estimates for EST were larger than their LCMCR counterparts. A key observation is that in most of these strata, it was not possible to fit a log-linear model complex enough to adequately model the joint distribution. We believe that the reason for this is that most of the dependency in the data is actually induced by individual heterogeneity, which the log-linear modeling strategy cannot not \textit{easily} account for---a clear advantage for LCMCR. One observable example of the extreme capture heterogeneity in the data is the difference in listability for victims of EST and SLU. This is the main reason why in this case we trust LCMCR estimates over the log-linear approach: log-linear models account for departures from independence exclusively through interactions terms between lists. This strategy can in principle approximate (or even directly represent) some configurations of heterogeneity-induced dependency \citep[see][]{Darroch1993,johndrow2017tensor}. However, in many cases the complexity of the joint distributions require models with a large number of interaction terms, including many high-order interactions. These complex log-linear models often run into identifiability problems due to sparsity of counts in the higher-order intersections, and the combinatoric explosion of possible log-linear models with more than a few lists makes them computationally infeasible. We encountered all these problems in our attempts to use log-linear models in our analyses. On the other hand, LCMCR works by directly modeling patterns of differential individual capturability, so for our data it is generally easier to find a parsimonious LCMCR representation than a log-linear one \citep[see also discussion in][]{Manrique:LCM:2016:Capture:Recapture}. We believe that if the situation were reversed (if the dependence were actually generated through list interactions), LCMCR could run into similar problems \citep[see][]{johndrow2017tensor} and log-linear would be more appropriate.

Our data are complex, and their analysis involved a great deal of exploration, modeling and decision-making. In this process we have learned some lessons that we believe can be useful for similar future projects. First, even with the new data, not all deaths in the conflict have been documented. Deaths tend to have highly variable probabilities of being observed. This suggests that we should prefer CR models that can address at least moderate capture heterogeneity, including log-linear models when the data are adequate to estimate high-order interactions. In this case, the sparse higher-order intersections in the data made LCMCR a better choice. Nonetheless, stratum sample size matters. Larger strata may contain elements with more heterogeneous capture probabilities. However their larger size can also allow to detect and account for more subtle structures; see \citet{Johndrow:Lum:Manrique:2016:CR:Heterogeneity} for an explanation of this phenomenon. Our alternative analysis using log-linear models suggest that comparisons among estimation approaches can illuminate data issues that do not admit direct testing, especially capture heterogeneity. Finally, we note that substantive guidance on stratification is essential to incorporate qualitative and historical information about the likely patterns of data capture.

Conflicts that involve governments, including civil wars and inter-state wars, tend to produce many victims. At the same time, these conflicts tend to degrade the social mechanisms that record vital statistics about mortality. Of course state registration systems break down, but media reports become more poorly sourced as journalists are targeted and conflict regions become less accessible; religious institutions themselves suffer attacks and become less able to record deaths; non-governmental monitoring organizations face logistical and security challenges collecting information. Most of all, witnesses and victims face the same logistical challenges, and they have good reason to fear retaliation if they denounce violence. Consequently, data about deaths in conflict is nearly always partial. Even in situations where more information becomes available as the conflict recedes into the past, only in a vanishingly small number of cases is complete information available (for one exception, see \cite{kruger2014evaluation}). Deaths in Peru have never been fully documented, and therefore, the only way the victims' stories can be told is by statistical estimation.

\section*{Acknowledgements}
The authors thank Eduardo Vega-Luna and Marco Pacheco from the Peruvian Ombudsman Office for their help allowing us access to the CVR data. We also thank Michelle Dukich for her assistance in the record-linkage phase. Text in Section~\ref{sec:conflict} adapts and reproduces small parts of \cite{schubiger:sulmont:peru:2019}.

\section*{Appendix A: Geographic Stratification Scheme}
Here we detail the geographic stratification scheme that we created for helping control the heterogeneity. We started with a 12-region scheme, and then sub-divided 5 of these to create a 26-region stratification. All decisions about grouping have been made based on qualitative information about the dynamics of the conflict.

Below we detail our scheme and explain its rationale.

\maketitle

\begin{itemize}
  \item \texttt{AYA\_CENT} (``Central Ayacucho''): Provinces of Huamanga, Cangallo, Vilcas-Huaman, Victor~Fajardo, Sucre, and Huanca~Sancos, in Ayacucho department. 

    Central Ayacucho was the first region of the country where the Shining Path initiated its armed insurrection. Along with North Ayacucho, this was the area of the country with the highest levels of casualties. At the beginning of the conflict, central Ayacucho was the main stronghold of the Shining Path and where this organization had heavily invested in political proselytism since the 1970's.

    We sub-divided this region into: 
    \begin{itemize} 
      \item \texttt{AYA\_CENT\_HUAMANGA\_AYA}:       District of Ayacucho (in Huamanga province).
      \item \texttt{AYA\_CENT\_CANGALLO\_CANGALLO}:  District of Ayacucho (in Cangallo province)..
      \item \texttt{AYA\_CENT\_VH\_VH}:              District of Vilcas~Huaman (in Vilcas~Huaman province).
      \item \texttt{AYA\_CENT\_HUAMANGA}: Rest of Huamanga province.
      \item \texttt{AYA\_CENT\_CANGALLO}: Rest of Cangallo province.
      \item \texttt{AYA\_CENT\_VH}: Rest of Vilcas~Huaman province.
      \item \texttt{AYA\_CENT\_VFSUCREHS}: Provinces of Victor~Fajardo, Sucre, and Huanca~Sancos.
    \end{itemize}

    This subdivision tries to distinguish the main urban centers from its most rural hinterland and the periphery, in order to capture the different dynamics of the conflict in those settings. The number of cases available in those areas facilitates this subdivision.

  \item \texttt{AYA\_NORTE} (``North Ayacucho''): Provinces of Huanta and La Mar, in Ayacucho department.

    North Ayacucho was the first region where the Shining Path expanded its activities. It is the region where all the projects have recorded the highest number of victims. 

    We sub-divided this region into:
    \begin{itemize}
      \item \texttt{AYA\_NORTE\_CAPITALES}: Huanta district (in Huanta province), and San~Miguel district (in La~Mar province).
      \item \texttt{AYA\_NORTE\_HUANTA}: Rest of Huanta Province. 
      \item \texttt{AYA\_NORTE\_LAMAR}: Rest of La~Mar province.
    \end{itemize}

The subdivisions in this region follows the same logic as in the previous region.

  \item \texttt{AYA\_NORTE\_CHUNGUI}: Chungui district (in Chungui province, Ayacucho department).

    Chungui in North Ayacucho, was the most affected district in the country during the armed conflict in terms of the proportion of recorded and estimated number of victims in relation to its population. It is also the district where we have one of the highest levels of coverage in term of the variety of sources available. 

  \item \texttt{AYA\_SUR} (``South Ayacucho''): Provinces of Lucanas, Parinacochas, and Paucar~Del~Sara~Sara, in Ayacucho department.

    These regions of Ayacucho were much less affected by the conflict than the rest of the department. Thus we can consider them peripheral areas, close to the central area of the country. 

 \item \texttt{SIERRA\_CENTRO} (``Central Highlands''): Departments of Pasco, Huancavelica, and the rest of Junin department.

    This region was the first region outside Ayacucho, where the Shining Path expanded its activities after the initial government crackdown against the Shining Path in Ayacucho between 1983-1984. 

    We sub-divided this region into:
    \begin{itemize}
      \item \texttt{SIERRA\_CENTRO\_PASCO}: Department of Pasco.
      \item \texttt{SIERRA\_CENTRO\_JUNIN}: Rest of department of Junin.
      \item \texttt{SIERRA\_CENTRO\_HUANCAVELICA}: department of Huancavelica.
    \end{itemize}
    
    This subdivision follows the main departments and provinces in the region affected with the conflict and where we can distinguish different dynamics and at the same time, have enough information to make them visible.
    
\item \texttt{SATIPO}: Satipo province in Junin department.

    Satipo is one of the most affected provinces in the central area or the country. In this area two of the insurgents groups (MRTA and Shining Path) had an important presence and particularly targeted and decimated the population of the Ashaninka ethnic group.

  \item \texttt{NOR\_ORIENTE} (``Northeast''): Departments of San~Martin and Huanuco; provinces Coronel~Portillo and Padre~Abad from Ucayali department.

    Those regions were more affected by the conflict by the second half of the 1980's and at the beginning of the 1990s. The armed conflict in those regions was also closely intertwined with drug production (peasants producers of coca leaves and cocaine raw materials) and drug trafficking. 

    We sub-divided this region into:
    \begin{itemize}
      \item \texttt{NOR\_ORIENTE\_SAN\_MARTIN}: Department of San~Martin.
      \item \texttt{NOR\_ORIENTE\_HUANUCO}: Department of Huanuco.
      \item \texttt{NOR\_ORIENTE\_UCAYALI}: Provinces of Coronel~Portillo and Padre~Abad (in Ucayali department).
    \end{itemize}

    This subdivision follows the main departments and provinces in the region affected with the conflict and where we can distinguish different dynamics and at the same time, have enough information to make them visible.

  \item  \texttt{SIERRA\_SUR} (``South Highlands''): Departments of Cusco, Puno, and Apurimac.

As in the case of the Central Highlands, the Shining Path expanded its activities in those regions by the mid 1980's. However, unlike the Central Highlands, it was less successful in mobilizing some communities or in organizing armed activities.

    We sub-divided this region into:
    \begin{itemize}
      \item \texttt{SIERRA\_SUR\_CUSCO}: Department of Cusco.
      \item \texttt{SIERRA\_SUR\_PUNO}: Department of Puno.
      \item \texttt{SIERRA\_SUR\_APURIMAC}: Department of Apurimac.
    \end{itemize}

This subdivision follows the main departments and provinces in the region affected with the conflict and where we can distinguish different dynamics and at the same time, have enough information to make them visible.

  \item \texttt{LIMA\_CALLAO}: Province of Lima (department of Lima), and Callao Constitutional province.

The main urban center of the country is considered a stratum by itself, not only because it concentrated almost a third of the country's population, but because of the symbolic and political repercussions of the armed activities and casualties that took place in this region.

  \item \texttt{LIMA\_PROVINCIAS} (``Provinces of Lima''): Rest of department of Lima.

    These provinces surrounded the main urban center of the country and they had a strategical value for subversive groups, which might explain the armed activity and pattern of victimization that took place in those areas. 

 \item \texttt{PERIFERIA} (``Surroundings''): Departments of Ica, Arequipa, Moquegua, Tacna, Ancash, La~Libertad, Cajamarca, Lambayeque, Piura, Tumbes, and Amazonas.

These regions were some of the least affected by the conflict. Mainly along the coast and in the north of the country. Insurgent groups have much less roots and presence and capabilities to organize armed activities. 

  \item \texttt{SELVA} (``Amazon Region''): Departments of Loreto, Madre~De~Dios, and the rest of Ucayali department.

    These regions were mainly peripheral areas of the conflict in the Amazon region, much less affected than the rest of the country.

\end{itemize}

\section*{Appendix B: An MCMC sampler for the LCMCR model with incomplete stratification}
\label{appen:LCMCR:strat:MCMC}

Starting from the model in \eqref{eq:LCMCR:strat}-\eqref{eq:LCMCR:strat:priors}, let us assume that any individual with $\bx_i = 0$ cannot be observed. Let us also assume that stratification labels for any individual with $i \in \cM \subset\{1..n\}$ is missing. Let $\mathcal{X} = \{\bx: \bx = \bx_i \text{ for some } i\leq n\}$ be the set of all observed capture patterns. We define the following random variables for $\bx \in \mathcal{X}$, $y \in \cY$ and $k = 1,...,K^*$:
\begin{itemize}
  \item Observed individuals from latent class $k$ with capture pattern $\bx$ and observed stratum labels of level $y$:
    $$\omega^{\bx,y}_{obs}(k) = \#\{i \notin \cM: y_i = y, \bx_i = \bx \neq \bzero, z_i = k\}$$
    Also, let $\omega^{\bx,y}_{obs} = \sum_k\omega^{\bx,y}_{obs}(k)$. We note that $\omega^{\bx,y}_{obs}$ is fully observed, while $\omega^{\bx,y}_{obs}(k)$ depends on unobserved $z_i$s.

  \item Observed individuals from latent class $k$ with capture pattern $\bx$ and missing stratum labels with true value $y$:
    $$\omega^{\bx,y}_{mis}(k) = \#\{i \in \cM: y_i = y, \bx_i = \bx \neq \bzero, z_i = k\}.$$
  Also, let $\omega^{\bx,y}_{mis} = \sum_k\omega^{\bx, y}_{mis}(k)$.

  \item Unobserved individuals from latent class $k$ with stratum labels with true value $y$.
    $$\omega^{\bzero, y}(k) = \#\{i : y_i = y, z_i = k, \bx_i = \bzero\}.$$
    Also, let $\omega^{\bzero,y} = \sum_k\omega^{\bzero,y}(k)$.
\end{itemize}
Let also $\bomega_{obs}$, $\bomega_{mis}$ and $\bomega^{\bzero}$ be tensors containing all the values of $\omega^{\bx,y}_{obs}(k)$, $\omega^{\bx,y}_{mis}(k)$ and $\omega^{\bzero,y}(k)$ respectively,  conveniently arranged. We note that for any given stratum $n_y = \sum_{\bx,y} \left(\omega^{\bx,y}_{mis} +  \omega^{\bx,y}_{obs}\right)$, and $N_y = n_y + \omega^{\bzero,y}$.

Assuming a prior distribution $p(N) \propto 1/N$ for $N = \sum_y N_y$, we can now write the posterior distribution of $\left(\bomega_{obs}, \bomega_{mis}, \bomega^\bzero, \boldsymbol{\alpha}, \blambda, \bpi, \brho\right)$ conditional on observed data $((\bx_i)_{i \leq n}, \by_{obs})$ only:
\begin{align}
  p&\left(\bomega_{obs}, \bomega_{mis}, \bomega^\bzero, \boldsymbol{\alpha}, \blambda, \bpi, \brho,\mid (\bx_i)_{i \leq n}, \by_{obs} \right) \\
  &\propto \dirichletd(\brho|1,..,1)  (N-1)!
  \prod_{y \in \cY}\Bigg\{
  SB_{K^*}(\bpi_k|\alpha_y) I(\omega^{\bzero,y} \geq 0)
    \prod_{k=1}^{K^*}\Bigg[
      \prod_{j=1}^J \betad(\lambda_{yjk}|p_y, q_y)\times \\
       & \qquad \frac{\left[\rho_y \pi_{yk}\prod_{j=1}^J(1-\lambda_{yjk})\right]^{\omega^{\bzero,y}(k)}}{\omega^{\bzero,y}(k)!}
      \times \prod_{\bx \in \mathcal{X}} \frac{\left(\rho_y \pi_{y,k}\prod_{j=1}^J\lambda_{yjk}^{x_{j}}(1- \lambda_{yjk})^{1-x_{j}}\right)^{\omega^{\bx,y}_{obs}(k) + \omega^{\bx,y}_{mis}(k)}}{{(\omega^{\bx,y}_{obs}(k) + \omega^{\bx,y}_{mis}(k))!}}
    \Bigg]
  \Bigg\}
\end{align}

From here we can easily derive a Gibbs sampler:
\begin{enumerate}
\item \textbf{Sample from $p(\bomega_{obs}|...)$ and $p(\bomega_{mis}|...)$}: For $\xi \in \{obs, mis\}$, $y\in \cY$, and $\bx \in \mathcal{X}$ sample
$$
\omega_\xi^{\bx,y}(k) \indep \multinomiald(\omega_\xi^{\bx,y}, (p_1,...,p_{K^*}))
$$
where $p_k \propto \pi_{yk} \prod_j\lambda_{yjk}^{x_j}(1-\lambda_{yjk})^{1-{x_j}}$

\item \textbf{Sample from $p(\boldsymbol{\lambda}|...)$}: For $y \in \cY$, $j=1, \dots , J$, $k=1, \dots , K^*$  let $m^1_{yjk} = \sum_{\bx \in \mathcal{X}} I(x_j=1)(\omega_{obs}^{\bx,y}(k) + \omega_{obs}^{\bx,y}(k))$ and $m^0_{yjk} = \omega^{\bzero,y}(k) + \sum_{\bx \in \mathcal{X}} I(x_j=0)(\omega_{obs}^{\bx,y}(k) + \omega_{obs}^{\bx,y}(k))$. Then sample
  $$\lambda_{yjk} \indep \text{Beta}\left( m^1_{yjk} +p_y, m^0_{yjk} + q_y \right).$$

\item \textbf{Sample from $p(\bpi_y|...)$}: For $y \in \cY$ and $k=1,...,K^*-1$ sample
  $$V_{yk} \sim \betad\left(1+ \nu_{yk}, \alpha_y + \sum_{h=k+1}^{K^*}\nu_{yh}\right)$$
    where $\nu_{yk} = \omega^{\bzero,y}(k) + \sum_{\bx}\left(\omega_{obs}^{\bx,y}(k) + \omega_{mis}^{\bx,y}(k)\right)$. Let $V_{yK^*}=1$ and make $\pi_{yk} = V_{yk}\prod_{h<k}(1-V_{yh})$ for all $k=1,...,K^*$.

\item \textbf{Sample from $p(\balpha|...)$}: $\alpha_y \sim \text{Gamma}(a - 1 + K^*, b - \log\pi_{yK^*})$

\item \textbf{Sample from $p(\boldsymbol{\omega}^{\bzero}|...)$}: The full joint conditional distribution of $\boldsymbol{\omega}^\bzero$ is
\begin{align}
p(\boldsymbol{\omega}^\bzero |...) \propto (N - 1)!
\prod_{y\in \cY}\prod_{k=1}^{K^*} \frac{\left[\rho_k \pi_{yk}\prod_{j=1}^J(1-\lambda_{yjk})\right]^{\omega^{\bzero,y}(k)}}{\omega^{\bzero,y}(k)!}
\end{align}
This expression is proportional to a negative multinomial distribution. A strategy for sampling from this distribution is
\begin{enumerate}
	\item Sample $n_0 \sim \text{NegBinomial} \left(n, 1 - \sum_{y \in \cY}\sum_{k=1}^{K^*} \rho_y\pi_k \prod_{j=1}^J(1-\lambda_{yjk}) \right)$.
	\item Sample $(\omega^{\bzero,y})_{y \in \cY} \sim \text{Multinomial}\left(n_0,(\eta_y)_{y \in \cY}\right)$ for $\eta_y \propto \sum_{k=1}^{K^*} \rho_y\pi_k \prod_{j=1}^J(1-\lambda_{yjk})$.
    \item For $y \in \cY$ sample $\left(\omega^{\bzero,y}(1),...,\omega^{\bzero,y}(K^*)\right) \sim \text{Multinomial}\left( \eta_k \right)$
    where $\eta_k \propto \pi_k \prod_{j=1}^J(1-\lambda_{yjk})$.
\end{enumerate}
\item \textbf{Sample from $p(\brho|...)$}: $\brho \sim \dirichletd\left( (N_y + 1)_{y \in \cY} \right)$
\end{enumerate}

\bibliography{./MSE_peru}

\end{document}